# Simulation-based Optimization of Toll Pricing in Large-Scale Urban Networks using the Network Fundamental Diagram: A Cross-Comparison of Methods

Ziyuan Gu, Meead Saberi*

*Research Center for Integrated Transport Innovation (rCITI), School of Civil and Environmental Engineering, University of New South Wales (UNSW), Sydney, Australia*

*\* Corresponding author: meead.saberi@unsw.edu.au*

**Abstract**

Simulation-based optimization (SO or SBO) has become increasingly important to address challenging transportation network design problems. In this paper, we propose to solve two toll pricing problems with different levels of complexity using the concept of the macroscopic or network fundamental diagram (MFD or NFD), where a large-scale simulation-based dynamic traffic assignment model of Melbourne, Australia is used. Four computationally efficient SBO methods are applied and compared, including the proportional-integral (PI) controller, regressing kriging (RK), DIviding RECTangles (DIRECT), and simultaneous perturbation stochastic approximation (SPSA). The comparison reveals that these methods work equally well on the simple problem without exhibiting significant performance differences. But, for the complex problem, RK manifests itself to be the best-performing method thanks to its capability of filtering out the numerical noise arising from computer simulations (i.e. allowing for non-smoothness of the objective function). While the PI controller is a more competitive solution to the simple problem given its faster rate of convergence, the poor scalability of the method in the complex problem results in limited applicability. Two caveats, however, deserve emphasis: (i) the chosen critical network density of the NFD does not necessarily represent a robust network control or optimization threshold, as it might shift in the presence of toll pricing; and (ii) re-interpolation is required as part of RK in order to achieve global convergence.

## 1. Introduction

Simulation-based optimization (SO or SBO) refers to, without loss of generality, the minimization of an objective function subject to a set of constraints, both of which are evaluated through stochastic simulations (Amaran et al., 2016). In transportation network design problems (NDPs), SBO is frequently applied whenever a "black-box" simulation model is present. Obtaining an analytical solution is impossible due to the lack of an explicit mathematical formulation of the traffic system in question, and so are any commonly used numerical solutions that require knowledge of the gradient or Hessian. Even when traffic dynamics is described by a mathematical model, e.g. a system of deterministic differential algebraic equations[1], rather than through a stochastic "black-box" simulator, one may still find it too difficult to calculate the gradient or Hessian and thus, derivative-free optimization (deterministic version of SBO) holds as a competitive solution (see Cheng et al. (2019b) and Zheng and Geroliminis (2016)).

This paper proposes two toll level problems (TLPs) with different levels of complexity using the macroscopic or network fundamental diagram (MFD or NFD). A large-scale simulation-based dynamic traffic assignment (SBDTA) model of Melbourne, Australia is used, which was previously calibrated and validated (Shafiei et al., 2018). We review, apply, and compare four state-of-the-art SBO methods as the solution algorithms to the TLPs, including the proportional-integral (PI) controller, regressing kriging (RK), DIviding RECTangles (DIRECT), and simultaneous perturbation stochastic approximation (SPSA). Among the four methods, RK and SPSA are intrinsically capable of dealing with simulation stochasticity, as is present in the SBDTA model. The PI controller and DIRECT are, in contrast, deterministic methods that belong to derivative-free optimization; but with the help of sample-path optimization, they can also be applied to stochastic problems (Amaran et al., 2016). By performing computational experiments and quantitative comparison of these methods on a large-scale real urban network (which have never been done in any previous studies), we offer new insights into better understanding, choosing, and implementing SBO when applied to solve toll pricing problems.

The remainder of the paper is organized as follows. Section 2 provides an overview of SBO and TLPs. Section 3 introduces the SBDTA model of Melbourne, Australia, based on which two TLPs are proposed. Section 4 details the four methods, which are applied and compared in Section 5. Section 6 concludes the paper.

## 2. Literature review

SBO is an active area of research in transportation science and engineering. It is typically applied to solve transportation NDPs when a traffic simulation model is used. There are a few studies reviewing different SBO methods purely from the algorithmic perspective; see Amaran et al. (2016); Figueira and Almada-Lobo (2014); Fu (1994, 2002); Rios and Sahinidis (2013). In general, SBO can be categorized into two broad categories, one with discrete variables (i.e. combinatorial optimization) and the other with continuous variables (i.e. continuous optimization). In this paper, we focus on the latter.

Continuous SBO methods can be further classified into seven categories: (i) random search, also known as metaheuristics; (ii) response surface methods (RSMs), also known as surrogate-based optimization or metamodeling; (iii) stochastic approximation (SA); (iv) direct search, also known as patten search; (v) estimation of distribution algorithms (EDAs); (vi) Lipschitzian optimization; and (vii) feedback control. As we mentioned in Section 1, some of them are, in fact, deterministic methods and thus, to be more accurate, belong to derivative-free optimization; but they can also be applied to solve SBO problems using sample-path optimization, which provides a deterministic realization of the stochastic function (Amaran et al., 2016). Thus, following the same reference, we present and discuss these methods all as SBO, although one must be aware of the formal (or most accurate) terminology.

---

[1] One shall distinguish between the numerical noise of a differential system resulting from finite-difference approximation and the numerical noise to be defined and emphasized in Section 4.

The applicability of a specific SBO category depends on the characteristics of the problem at hand. A TLP coupled with a large-scale SBDTA model often features an expensive-to-evaluate objective function, multiple toll decision variables, and simulation stochasticity (Chen et al., 2014). The latter is typically realized by setting different random seed numbers in the traffic simulator, which must not be confused with the numerical noise of computer simulations (i.e. non-smoothness of the objective function) to be introduced and demonstrated in Sections 4 and 5, respectively. To find the optimal solution in a computationally efficient manner, a "smart" enough SBO category must be employed requiring fewer objective function evaluations. Thus, metaheuristics (see Elbeltagi et al. (2005) for an overview and comparison) and EDAs (see Kroese et al. (2006) for the cross-entropy method) are not advisable given their demanding requirements of many objective function evaluations that lead to a greater computational burden (Amaran et al., 2016). Chow et al. (2010) and Chen et al. (2014) compared RSMs with the genetic algorithm (GA), showing that the former converged much faster than the latter. More recently, Osorio and Punzo (2019) formulated an RSM that incorporated the problem-specific knowledge from an analytical traffic model, showing that it outperformed SPSA and pattern search.

There are many successful applications of SBO to a variety of transportation problems, such as model calibration (Cheng et al., 2019b; Hale et al., 2015; Lu et al., 2015; Osorio, 2019; Osorio and Punzo, 2019; Sundaram et al., 2011; Zhang et al., 2017), signal optimization (Chong and Osorio, 2017; Osorio and Bierlaire, 2013; Osorio and Chong, 2015; Osorio and Nanduri, 2015; Zheng et al., 2019), and bus transit operations (Wu et al., 2019; Zhang and Xu, 2017). One growing application is to toll pricing, a widely discussed travel demand management strategy to address the increasing traffic congestion; see Gu et al. (2018a) for an overview of different pricing practices in multiple cities or regions. Mathematically speaking, a TLP is typically formulated as a bi-level optimization problem (Yang and Huang, 2005), where the upper level forms the objective function(s) to be minimized or maximized (from the government's perspective) and the lower level consists of a traffic assignment model representing users' collective behavior in response to pricing. The objective of toll pricing can be multiple including but not limited to total travel time minimization (Chen et al., 2014), network speed regularization (Liu et al., 2013), total revenue maximization (Chen et al., 2016), network travel time reliability maximization (Chen et al., 2018), and network flow maximization (Zheng et al., 2016). Each objective corresponds to a unique perspective from which the network is evaluated, and thus different researchers and practitioners may have different preferences. To solve TLPs, previous studies generally employed feedback control (Gu and Saberi, 2019; Gu et al., 2018c; Simoni et al., 2015; Zheng et al., 2016; Zheng et al., 2012) or RSMs (Chen et al., 2016; Chen et al., 2014; Chen et al., 2018; Chow and Regan, 2014;

Chow et al., 2010; Ekström et al., 2016; He et al., 2017; Rodriguez-Roman and Ritchie, 2018) as the solution algorithms.

There is limited research on the application and comparison of different SBO methods for solving transportation NDPs. Huang et al. (2006) is one of the few studies that applied and compared several methods, but mainly on noisy test functions rather than on transportation problems. While some other studies (as we previously discussed) applied a certain method to a certain transportation problem, they fall short in providing adequate discussion on how and why they chose the methods in the first place, and whether some other methods might behave similarly or differently. This is the motivation behind this study, where we endeavor to solve two toll pricing problems on a large-scale real urban network using and comparing different SBO methods. While the quantitative comparison is not intended to be a comprehensive guide to the application of these methods accounting for every possible transportation problem, it offers new computational and practical insights into better understanding, choosing, and implementing them when applied to solve toll pricing problems.

## 3. Problem formulation

We first briefly introduce the SBDTA model of Melbourne, Australia in Subsection 3.1. The model is part of the formulation of the TLPs to be described in Subsection 3.2.

### 3.1. Simulation model

Deployed in AIMSUN, the mesoscopic SBDTA model of Melbourne, Australia was calibrated and validated using multi-source traffic data. The model is open sourced which can be accessed via https://github.com/meeadsaberi/dynamel. Related documentation and references can be found in Gu et al. (2016, 2018b); Shafiei et al. (2018); Shafiei et al. (2017). Fig. 1a shows the extracted sub-network from the greater Melbourne area model that is used in this paper. Bounded by the red dash lines, the sub-network has a total of 4,375 links, 1,977 nodes, and 492 centroids. The inner rectangle represents the pricing zone (PZ) covering the Melbourne CBD where traffic congestion is usually severest during the rush hours. There are in total 282 links, 91 nodes, and 30 centroids in the PZ. The simulation covers the 6-10 AM peak period with calibrated origin-destination (OD) demand from historical data for every 15 minutes. Accordingly, path assignment (with or without toll pricing) is performed every 15 minutes using the C-logit model (Cascetta et al., 1996) with default parameter settings. To achieve a more realistic initial network state, a 30-min warm-up period is introduced at the beginning of the simulation. Another 30-min zero-demand period is added to the end of the simulation to help recover the network. We assume the average value of time is 15 $/h (Legaspi and Douglas, 2015) and 30 percent of drivers are adaptive; i.e., 30 percent of drivers who remain in the network at the new path assignment interval (who are randomly drawn irrespective of their OD pairs) would re-evaluate their current assigned shortest paths and update them if any new shorter paths are available given the prevailing traffic conditions in the

network[2]. Note, however, that 30 percent does not necessarily represent a realistic proportion in practice; this value is chosen based on the documentation (Shafiei et al., 2018) as well as Saberi et al. (2014b) who provided an in-depth analysis of adaptive driving in a large-scale urban network.

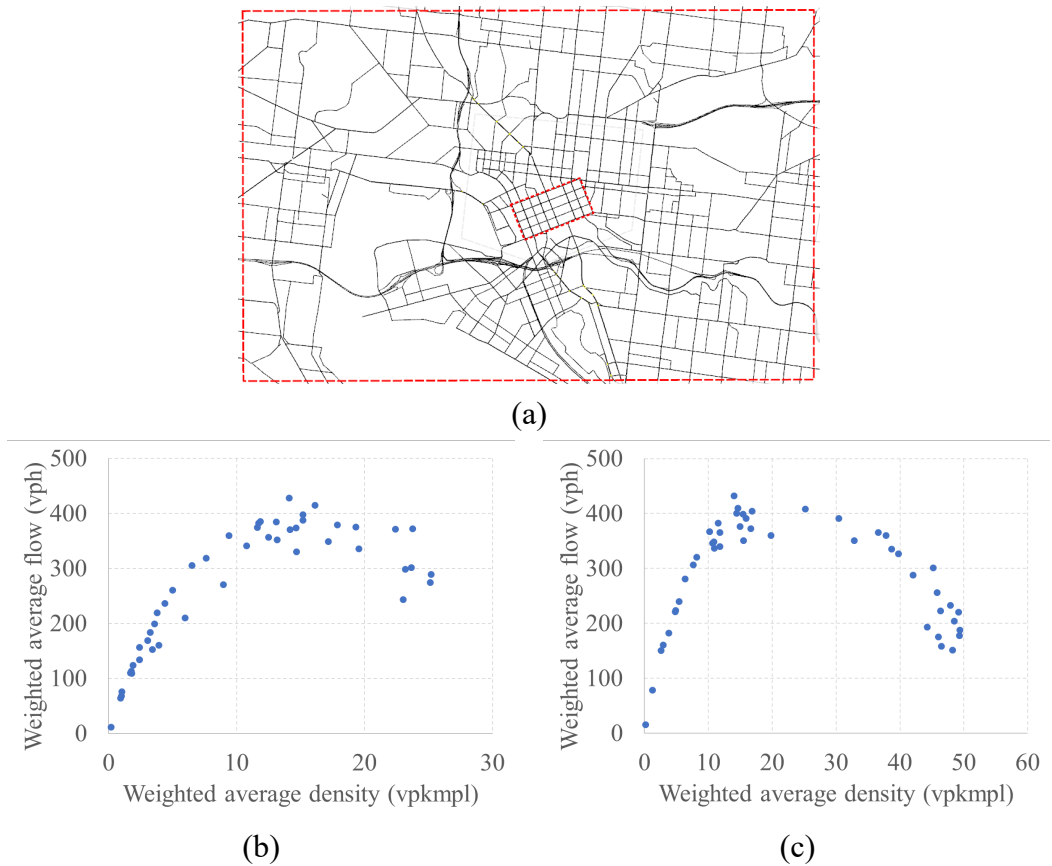

**Fig. 1.** (a) Extracted sub-network from the greater Melbourne area model; and simulated NFDs under the non-tolling scenario for (b) the simple problem and (c) the complex problem.

*3.2. Toll level problems*

In this paper, the objective of TLPs is chosen to maximize network flows, which has recently received increasing attention due to the re-theorization of the NFD (Geroliminis and Daganzo, 2008). The logic (from the same reference) is that the network flow has a robust linear relation with the trip completion rate provided a relatively constant average trip length, which suggests that a larger network flow equates to a better network productivity. Since the critical network density of the NFD is known to correspond to the maximum network flow, the objective function then aims to minimize the differences or deviations between the observed/simulated network densities and this critical value. To estimate the critical network density, one can inspect the observed/simulated NFD and choose the abscissa of the maximum network flow (Zheng et al., 2012), or fit a function to the NFD observations and use its derivative (Zheng et al., 2016). Note, however, that an (estimated) NFD function is not required in this work because the NFD is used only as an indicator of the network congestion level rather than being fed into any mathematical models. Thus, the natural existence of the scatter and hysteresis in the NFD is not a major concern. Nevertheless, one critical assumption is that the shape of the NFD including the position of the critical network density does not change significantly in the presence of toll pricing; otherwise, the critical network density is a moving target that cannot be effectively used for network control or optimization purposes. While this assumption was commonly made in many previous studies (Gu et al., 2018c; Gu et al., 2019; Zheng et al., 2016; Zheng et al., 2012), it is not guaranteed to hold under all possible circumstances (see Subsection 5.2). In fact, violation of this assumption was demonstrated already using signal control (Alonso et al., 2019; Keyvan-Ekbatani et al., 2016; Zhang et al., 2013). In view of this potential limitation, we further formulate another objective function for comparison purposes which directly maximizes the observed/simulated network flows without relying on the specification of the critical network density. The resulting problem is still network flow maximization but without the need for identifying the NFD.

---
[2] The shortest paths determined from previous assignment intervals may no longer be the shortest ones due to the time-varying traffic conditions in the network.

While the TLPs to be formulated appear simple in structure, they involve a few complications: (i) employing an SBDTA model as a "black-box" function of the system dynamics prevents the application of any exact gradient methods; (ii) computer simulations may display a specific type of numerical noise – the objective function evaluations tend to scatter about a smooth trend rather than lying on it (Forrester et al., 2006) – which is likely to reduce the effectiveness and efficiency of certain SBO methods; and (iii) optimization using a large-scale SBDTA model typically leads to an expensive-to-evaluate objective function (Chen et al., 2014), thus requiring computationally efficient methods to avoid excessive objective function evaluations (i.e. simulation runs). Note that in the problem formulation, we only focus on optimizing the traffic conditions within the PZ without accounting for the peripheral network, same as Gu et al. (2018c); Gu et al. (2019); Simoni et al. (2015); Zheng et al. (2016); Zheng et al. (2012). The reason is twofold (Gu et al., 2018c): (i) city ring roads are often available around the city center providing sufficient road capacity for detour traffic (e.g. the M1, M2, and M3 highways surrounding the Melbourne CBD in Fig. 1); and (ii) to achieve a congested periphery, the OD demand has to climb to a level that is seldom realistic for normal daily traffic (e.g. a 15 percent increase at least for the Melbourne sub-network). The same reference showed that the Melbourne sub-network did not suffer from peripheral congestion, from the NFD perspective (not on an individual link basis), when the city center was optimally tolled. Nevertheless, if one pursues a systematic approach that explicitly accounts for peripheral congestion, further research efforts are required towards a coordinated modeling framework.

For better readability, a summary of notations is provided in Table 1.

**Table 1**

Summary of notations for problem formulations.

| Notation | Interpretation |
|---|---|
| $K_{cr}$ | Critical network density identified from the NFD |
| $\bar{K}_h/\bar{Q}_h$ | Average network density/flow during the $h$th tolling interval |
| $\bar{\mathbf{K}}$ | Vector of the average network densities for all the tolling intervals |
| $\boldsymbol{\tau}$ | Toll rate decision vector |
| $DTA(\boldsymbol{\tau})$ | "Black-box" function of the SBDTA model linking the input $\boldsymbol{\tau}$ to any output of interest |
| $\boldsymbol{\tau}_{\min}/\boldsymbol{\tau}_{\max}$ | Lower/upper bound of $\boldsymbol{\tau}$ |
| $\Omega$ | Feasible set of $\boldsymbol{\tau}$ |
| $h$ | Rank of the tolling interval |
| $m$ | Total number of tolling intervals |
| $\eta_h/\omega_h$ | Distance/delay toll rate for the $h$th tolling interval |
| $\alpha/\beta$ | Toll pattern smoothing parameter for $\eta_h/\omega_h$ |

*3.2.1. A simple problem*

To determine the critical network density $K_{cr}$, we run the simulation without pricing and show the simulated NFD of the PZ in Fig. 1b for a single run (multiple runs exhibit similar patterns). We follow Mahmassani et al. (1984) to calculate the simulated NFD; i.e., the average network flow, density, and speed are calculated as distance-weighted averages of link-based measurements, where distance is defined as the total lane kilometers of a link. The math is simple which can be easily found in the NFD literature, such as Saberi et al. (2014a), and thus is not repeated. For an overview and comparison of different methods to estimate the NFD, we refer to Leclercq et al. (2014).

The simulated NFD suggests that we set $K_{cr}$ at 15 vpkmpl[3] where the network flow reaches the maximum. This is consistent with the "AB rule" proposed by Daganzo (2007) – if the maximum network flow is achieved for a range of densities, any density within this range could be chosen as the critical threshold for network control purposes. One could opt for a smaller density to facilitate the stability of the system around this critical point (Zheng et al., 2016), which naturally leads to a longer tolling horizon and possibly a higher toll level. The chosen critical network density determines a 1-h tolling horizon between 8 AM and 9 AM (i.e. the peak period during which densities exceed $K_{cr}$). Without loss of generality, we set the length of individual tolling intervals at 30 minutes in view of the practice in Singapore (de Palma and Lindsey, 2011). Thus, we have a total of two intervals and a problem of two dimensions.

The dimension of a toll pricing problem is usually limited because (i) the tolling horizon typically covers peak hours only without considering inter-peaks when congestion is not severe; and (ii) from both the system's and user's perspectives, the length of individual tolling intervals is seldom set at a small value (Gu et al., 2018a). Note that such a two-dimensional setting largely facilitates the presentation and visualization of the numerical results, the comparison of the search behavior of different SBO methods, and the discussion on a few caveats including the effects of the numerical noise in computer simulations and the significance of re-interpolation in RK (see Subsection 5.1). Also note that the feedback control method, which was previously employed in Gu and Saberi (2019); Gu et al. (2018c); Zheng et al. (2016); Zheng et al. (2012), is only applicable to the simple problem due to its inability to consider many decision variables and complex constraints.

---

[3] Short for vehicles per kilometer per lane.

Under the assumption that the shape of the NFD including $K_{cr}$ does not change significantly when the network is pricing-controlled[4], we formulate the simple TLP considering a distance only toll:

$$\min_{\boldsymbol{\tau} \in \Omega} \mathbb{E}\left[\frac{1}{m}\sum_{h=1}^{m}|\bar{K}_h - K_{cr}|\right], \tag{1}$$

s.t.
$$\bar{\mathbf{K}} = DTA(\boldsymbol{\tau}), \tag{2}$$
$$\Omega = \{\boldsymbol{\tau}|\boldsymbol{\tau}_{\min} \leq \boldsymbol{\tau} \leq \boldsymbol{\tau}_{\max}\}, \tag{3}$$

where $\boldsymbol{\tau} = [\tau_1, ..., \tau_h, ..., \tau_m]^T$ and $\bar{\mathbf{K}} = [\bar{K}_1, ..., \bar{K}_h, ..., \bar{K}_m]^T$ are the decision vector of the distance toll rates and the vector of the average network densities for the $m = 2$ tolling intervals, respectively, $\mathbb{E}[\cdot]$ is the expectation operator, $DTA(\cdot)$ is the "black-box" function of the SBDTA model linking the toll price (i.e. simulation input) to the simulated average network densities (i.e. simulation output)[5], and $\Omega$ is the feasible set of toll rates with $\boldsymbol{\tau}_{\min} = [0,0]^T$ and $\boldsymbol{\tau}_{\max} = [1,1]^T$ being the lower and upper bounds, respectively. The objective function Eq. (1) minimizes the expected differences between the observed/simulated densities and $K_{cr}$ such that the network is driven to and operate in the capacity regime on the NFD plane.

We take the mathematical expectation in light of sample-path optimization, which allows deterministic methods to be applied to stochastic problems. It can also be incorporated with stochastic methods if simulation stochasticity is high, so as to reduce the effects of noise (Spall, 1998); e.g., three repetitions were took and averaged in Chen et al. (2018); Gu et al. (2019); He et al. (2017). In contrast, if simulation stochasticity is low or negligible and/or the computational time is of concern, one could opt for one replication only as in Chen et al. (2016); Chen et al. (2014); Chen et al. (2015); Ekström et al. (2016). Problem (1)-(3) represents an indirect flow maximization approach that solely replies on information of densities (including $K_{cr}$). The toll price for a trip is linearly dependent on the total distance traveled within the PZ, representing a usage-based pricing mechanism as opposed to the distance-independent pay-per-entry scheme. The latter might trigger concerns about inefficiency and inequity (Meng et al., 2012).

### 3.2.2. A complex problem

The simple problem has limited applicability because only two decision variables and bound constraints are present. Thus, we further formulate another complex problem involving more decision variables and generic constraints, so as to shed some light on the scalability of different methods in solving toll pricing problems. However, due to a higher dimensional space, the visualization of the numerical results naturally becomes less straightforward[6]. Under an increased demand scenario, the new simulated NFD of the PZ (see Fig. 1c) enters the congested regime without experiencing network recovery. We set $K_{cr}$ at 25 vpkmpl and obtain a 2-h tolling horizon between 8 AM and 10 AM. The length of individual tolling intervals is reduced to 15 minutes simply to increase the dimension of the toll rate decision vector, resulting in a total of eight intervals. Also, we introduce an arguably more efficient and equitable joint distance and delay toll into the complex problem, as opposed to the distance only toll in the simple problem. The problem dimension thus increases to 16 vs. the previous two. This rather new pricing mechanism was recently proposed and applied in Cheng et al. (2019a); Gu et al. (2018c); Gu et al. (2019) as a variant of Liu et al. (2014). We assume both the distance and the delay toll components are linear with respect to the distance traveled and the delay experienced within the PZ, respectively, where delay is defined as the difference between the observed/simulated travel time and the free-flow travel time. The formulation of the joint toll is simple which can be easily found in the above references; thus, it is not repeated.

---

[4] This assumption will be examined in Subsection 5.2.

[5] Thus, it represents the simulation input-output mapping.

While network flows maintain at the maximum for a range of densities from 20 to 30 vpkmpl (see Fig. 1c), we choose the center of this range as $K_{cr}$ and formulate the complex TLP:

$$\min_{\boldsymbol{\tau} \in \Omega} \mathbb{E}\left[\frac{1}{m}\sum_{h=1}^{m}|\bar{K}_h - K_{cr}|\right], \quad (4)$$

s.t.

$$|\eta_h - \eta_{h+1}| \leq \alpha, \ h = 1,2,\ldots,m, \quad (5)$$
$$|\omega_h - \omega_{h+1}| \leq \beta, \ h = 1,2,\ldots,m, \quad (6)$$
$$\bar{\mathbf{K}} = DTA(\boldsymbol{\tau}), \quad (7)$$
$$\Omega = \{\boldsymbol{\tau}|\boldsymbol{\tau}_{\min} \leq \boldsymbol{\tau} \leq \boldsymbol{\tau}_{\max}\}, \quad (8)$$

where $\boldsymbol{\tau} = [\eta_1, \ldots, \eta_m, \omega_1, \ldots, \omega_m]^T$ is the extended decision vector of toll rates for the $m = 8$ tolling intervals with $\eta_h$ and $\omega_h$ being the distance and the delay toll rates for the $h$th tolling interval, respectively, and $\alpha$ and $\beta$ are the associated toll pattern smoothing parameters. Constraints (5) and (6), also known as smoothing control constraints (Geroliminis et al., 2013), are used to ensure that the optimal toll rates do not fluctuate unduly between adjacent tolling intervals; thus, a smooth optimal toll pricing profile can be obtained. Without loss of generality, we set $\boldsymbol{\tau}_{\min} = [0,\ldots,0,0,\ldots,0]^T$, $\boldsymbol{\tau}_{\max} = [1,,\ldots,1,15,\ldots,15]^T$, $\alpha = \frac{1}{3}(1-0) \approx 0.33$, and $\beta = \frac{1}{3}(15-0) = 5$. A sensitivity analysis on $\alpha$ and $\beta$ can be found in Gu et al. (2019) showing that smaller values lead to a smoother toll pricing profile, as expected. This result is consistent with Geroliminis et al. (2013).

---

[6] One could employ certain projection methods for visualization in a low dimensional space; e.g., see Chen et al. (2018).

We proceed to discuss the potential limitation of the objective function Eq. (4). If the NFD exhibits a trapezoidal shape rather than a triangular one, as shown in Fig. 1c and analyzed in Daganzo (2007) and Mahmassani et al. (2013), $K_{cr}$ becomes somewhat ambiguous and no universal law exists as to the center of the plateau on the trapezoidal NFD plane being a good threshold. Further, if the assumption fails that the shape of the NFD does not change significantly when the network is pricing-controlled, the optimality of the problem solution cannot be guaranteed. To demonstrate this potential limitation, we now look directly at maximizing network flows throughout the tolling horizon, instead of identifying $K_{cr}$ around which the network shall operate. This can be achieved by simply replacing Eq. (4) with $\max_{\boldsymbol{\tau} \in \Omega} \mathbb{E}\left[\frac{1}{m}\sum_{h=1}^{m} \bar{Q}_h\right]$ where $\bar{Q}_h$ is the average network flow within the $h$th tolling interval, and $\bar{\mathbf{K}}$ in Eq. (7) with $\bar{\mathbf{Q}} = [\bar{Q}_1, \ldots, \bar{Q}_h, \ldots, \bar{Q}_m]^{\mathrm{T}}$ which is the vector of the average network flows for all the tolling intervals. The resulting toll pricing problem is intended to achieve the same effects as problem (4)-(8), but without the need for identifying the NFD. A comparison between the two is made in Subsection 5.2.1.

## 4. Solution algorithms: simulation-based optimization

To solve the two TLPs, we choose and apply four state-of-the-art SBO methods in view of their computational efficiency and global exploration. Instead of restricting ourselves to a specific method, we endeavor to provide a cross-comparison of multiple methods so that insights can be gained into better understanding, choosing, and implementing them when applied to solve toll pricing problems. These methods, including the PI controller, RK, DIRECT and SPSA, have been widely used in the last decade to solve different toll pricing problems. In Table 2, we provide a brief summary of the core of each method together with some of its applications in the recent literature. We emphasize, however, that the applications of these methods are not restricted to toll pricing problems only (see Section 2), but we prefer not to put the other references in the same table given the scope of the paper. One immediate observation consistent with our discussion in Section 2 is that the PI controller and RK enjoy greater popularity than the other two methods. In Section 5, we provide some quantitative evidence and explanation.

**Table 2**

Mechanisms and applications of the four SBO methods to toll pricing problems.

| Method | Mechanism | Applications |
|---|---|---|
| PI controller | To approach the set point via trial and error | Gu and Saberi (2019); Gu et al. (2018c); Zheng et al. (2016) |
| RK | To approximate the simulation input-output mapping by a mathematical construct | Chen et al. (2016); Chen et al. (2014); Chen et al. (2018); Gu et al. (2019); He et al. (2017) |
| DIRECT | To iteratively partition the search space where the fittest subspace survives | Gu et al. (2019); Hellman (2010) |
| SPSA | To approximate the gradient by finite difference | Xu (2009) |

Our problems are contaminated with simulation noise of a stochastic traffic simulator rendered by different random seed numbers. One could apply fixed-number sample-path optimization, also known as sample average approximation (Amaran et al., 2016), whose logic is straightforward – the sample mean is an unbiased estimator of the population mean; the larger the sample size, the better the approximation. One could also apply the "smarter" variable-number sample-path optimization (Deng and Ferris, 2007; Deng and Ferris, 2009), where the sample size varies for different decision vectors, as a more efficient way to handle simulation stochasticity. But the purpose is the same – we want to either make deterministic methods applicable to stochastic problems or reduce the effects of simulation stochasticity on stochastic methods. To measure the level of simulation stochasticity, one could run a few replications under certain scenarios and check statistics of the objective function values such as the coefficient of variation. It turns out that our Melbourne model exhibits rather low simulation stochasticity and thus, also to reduce the computational time, we follow Chen et al. (2016); Chen et al. (2014); Chen et al. (2015); Ekström et al. (2016) to perform one replication only.

The above simulation stochasticity injects noise into the optimization, but it is not the only source of noise that one might encounter. In fact, many computer simulations display what is termed the numerical noise (Forrester et al., 2006) – the objective function evaluations tend to scatter about a smooth trend rather than lying on it – which could reduce the effectiveness and efficiency of certain SBO methods (see Section 5). We emphasize that one must not confuse him/herself over this particular type of numerical noise vs. the previous noise arisen from simulation stochasticity. Even if the latter is negligible or otherwise is reduced by averaging across multiple replications, the former could still be present resulting in notable fluctuations of the (mean) trend; i.e., the response surface (RS) could be non-smooth even in the case of a deterministic model (e.g., see Cheng et al. (2019b)). Clearly, this numerical noise must not be captured by the optimization, which might (temporarily) mislead the search towards the global optimum or optima.

In the rest of this section, we briefly revisit each of the four SBO methods whose algorithmic details as well as notations are provided in Appendix A.

*4.1. Proportional-integral controller*

The PI controller is a classical feedback control method that has been widely used for traffic control and management. Example applications include ramp metering (Papageorgiou et al., 1991), perimeter or gating control (Keyvan-Ekbatani et al., 2012; Ramezani et al., 2015), and toll pricing (Gu and Saberi, 2019; Gu et al., 2018c; Zheng et al., 2016). The core of the method is simple – given a set point for an output of interest, the PI controller strives to minimize the discrepancies between the output and this set point by iteratively adjusting the input (i.e. the decision vector). Specifically, for toll pricing using the NFD, it works by iteratively adjusting the toll rates so that the NFD of the PZ maintains in the free-flow and capacity regimes. The detailed algorithmic steps including the formulation of the PI controller are provided in Appendix A.1.

*4.2. Regressing kriging*

Kriging originates from geostatistics but has become popular in designing and analyzing computer experiments since Sacks et al. (1989). It constructs, using an initial set of sample points, a coarse RS which is refined further in an iterative manner by adding infill sample points (e.g. via expected improvement (EI) sampling). Thus, the method works by learning and approximating the simulation input-output mapping, and can balance between global exploration and local exploitation (Jones et al., 1998). The workflow of RK with detailed description of each component is provided in Appendix A.2.

*4.3. DIviding RECTangles*

Lipschitzian optimization (Shubert, 1972) is a competitive method for global optimization, because (i) only the Lipschitz constant needs to be specified resulting in negligible effort for parameter tuning; and (ii) it can generate a lower bound on the optimal objective function value enabling the adoption of more meaningful stopping criteria. Nevertheless, limitations of the method also exist that prevent its further application: (i) the Lipschitz constant can be difficult to calculate or simply does not exist; (ii) it has a slow rate of convergence; and (iii) the computational efficiency decreases as the problem dimension increases. In view of these limitations, Jones et al. (1993) proposed a variant method termed DIRECT that eliminates the need for a user-specified Lipschitz constant. Equally contributive is that the method only samples and evaluates the midpoint of each partitioned hyperrectangle from the search space irrespective of the problem dimension, which largely reduces the number of objective function evaluations and thus the computational time. In a nutshell, DIRECT works by iteratively partitioning the search space into multiple hyperrectangles and identifying the potentially optimal ones for further partitioning. The detailed algorithmic steps are provided in Appendix A.3.

*4.4. Simultaneous perturbation stochastic approximation*

SA is a widely used method for solving challenging optimization problems that do not have an analytical solution. This happens when computer simulations are used, and thus the gradient is inaccessible which can only be approximated by direct objective function evaluations. While the two-sided finite-difference SA (Kiefer and Wolfowitz, 1952) is one such method, it is inefficient for solving a high dimensional problem as the number of objective function evaluations per iteration is twice the number of the problem dimension. In contrast, random direction SA or SPSA as a variant provides a highly efficient approximation to the gradient that requires only two objective function evaluations per iteration irrespective of the problem dimension (Spall, 1992). Due to its perturbation characteristic that offers similar statistical functionality to the Monte Carlo noise, SPSA can often work as a global optimizer (Spall, 2003). The detailed algorithmic steps including the formulation of the gradient approximation are provided in Appendix A.4.

## 5. Numerical results, comparison, and discussions

In this section, we first apply the four SBO methods to solve the simple TLP. Since the problem is two-dimensional, we can readily visualize and observe the performance of individual methods in searching for the global optimum or optima. Three of the four methods are then applied to solve the complex problem, except for the PI controller due to its inability to handle generic constraints.

*5.1. Solving the simple problem*

*5.1.1. Proportional-integral controller*
While effort is needed to tune the controller gain parameters, the PI controller does not require multiple runs as no random component is involved in the method. A rule of thumb is to start with a large pair of $P_P$ and $P_I$ and gradually reduce their values until a relatively smooth convergence pattern is achieved (Gu et al., 2018c). This is illustrated in Fig. 2. When we set $P_P = 0.02$ and $P_I = 0.005$, the resulting convergence pattern is smooth without significant oscillatory behavior; but when we set $P_P = 0.1$ and $P_I = 0.03$, the adjusted toll rates undergo radical changes between successive iterations and thus, one can hardly pinpoint the convergent solution especially for the second tolling interval. While making a wild guess is possible by referring to the center lines lying amid the fluctuations, the accuracy or solution quality is not guaranteed. Such a comparison highlights the importance of parameter tuning in the PI controller.

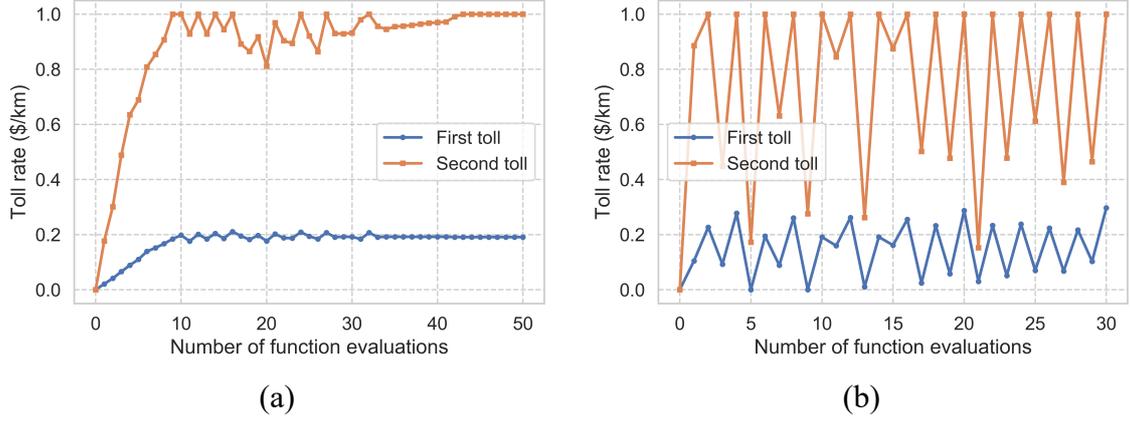

**Fig. 2.** Effects of different controller gain parameters on the performance of the PI controller: (a) $P_P = 0.02$ and $P_I = 0.005$; and (b) $P_P = 0.1$ and $P_I = 0.03$.

Fig. 3a shows the average network densities of the PZ during both tolling intervals, which gradually reduce to and stabilize around 15 vpkmpl (i.e. the pricing-control threshold $K_{cr}$) as the number of objective function evaluations increases to 50. The objective function evaluation decreases to the ideal optimum of zero despite exhibiting some fluctuations in the presence of the numerical noise. We observe that the toll rates quickly converge to their respective optima within about 10 objective function evaluations and only undergo minor changes afterwards (see Fig. 2a). This is somewhat contradictory to Fig. 3a where the fluctuations in the objective function value do not disappear for at least 30 objective function evaluations, and to Fig. 3b where the current best solution keeps reducing. There is still improvement in the objective function value between the 10$^{th}$ and 30$^{th}$ objective function evaluations, but at the same time the adjusted toll rates vary little. This is further illustrated in Fig. 3c by showing the search path of the PI controller. As expected, it quickly orients towards the optimal region which, however, covers a range of objective function values despite being relatively small in space (i.e. the numerical noise). Note that the PI controller works independently in each tolling interval; i.e., for a specific tolling interval, the input measurements to the PI controller only come from that interval without consideration for those from the other intervals. Thus, there is no interaction or coordination among different intervals (Gu et al., 2018c). This nature of the method (which explains why it can hardly be applied to many intervals) and the presence of the numerical noise jointly explain the existence of the fluctuations in Fig. 2a during the second tolling interval.

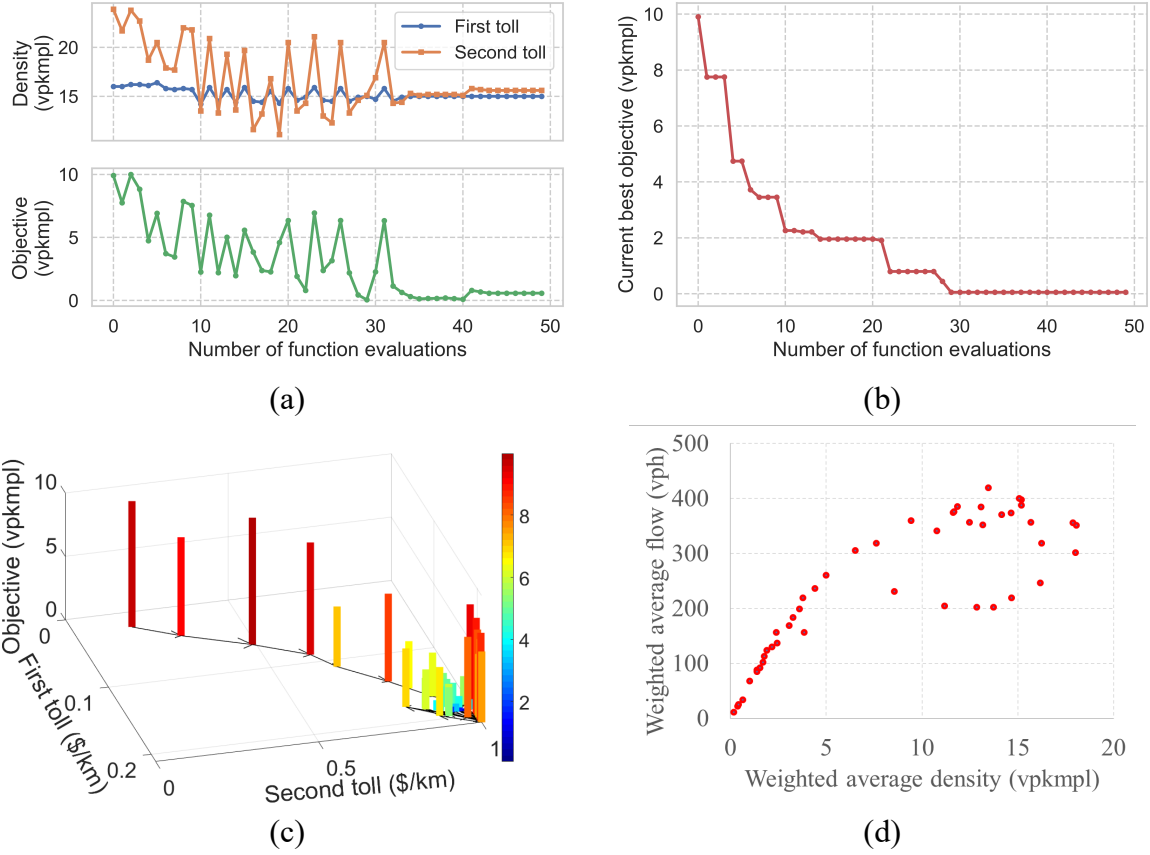

**Fig. 3.** Results obtained when applying the PI controller to solve the simple problem: (a) evolution of the network densities and the objective function value as the number of objective function evaluations increases; (b) evolution of the current best solution as the number of objective function evaluations increases; (c) search path of the method; and (d) simulated NFD of the PZ under the optimal tolling scenario.

By tracking the current best solution throughout the iterations, the optimal toll rates are 0.19 $/km for the first tolling interval and 0.93 $/km for the second. Fig. 3d shows that the PZ, after the optimal toll rates are applied, successfully operates around $K_{cr}$ without entering the congested regime that appears under the non-tolling scenario. There is, however, a notable hysteresis loop in the NFD, for which the reason is threefold: (i) hysteresis is an effect of network instability that naturally emerges during network recovery, when congestion tends more strongly towards unevenness as very congested areas clear more slowly than less congested ones (Daganzo, 2011; Gayah and Daganzo, 2011); (ii) the much higher toll price for the second tolling interval results in reduced inflow or demand to the PZ, thus amplifying the hysteresis loop (Gu et al., 2019)[7]; and (iii) the distance only toll per se could add to the significance of hysteresis, because drivers tend to accumulate themselves into the shortest paths within the PZ (which give the least toll price) resulting in a more heterogeneous distribution of congestion and thus a larger hysteresis loop in the NFD (Gu et al., 2018c).

---

[7] With demand dropping sharply to zero as an extreme case, the hysteresis loop in the NFD is amplified most significantly (Mahmassani et al., 2013).

*5.1.2. Regressing kriging*

RK requires the maximum likelihood estimations (MLEs) of several parameters throughout the iterations. Given that an initial set of sample points is required through random sampling, and that the GA is used to find both the MLEs and the optimal infill sample points, we perform multiple runs to consider this randomness effect. The probabilistic EI metric is calculated and maximized at each iteration to generate an optimal infill sample point, which does not apply to the initial sample points that comprise the first 11 objective function evaluations.

Fig. 4 shows the results for one run of RK while the results for other runs are provided in Appendix B. We can observe the expected decreasing trend in the probabilistic EI metric as the number of infill objective function evaluations increases (see Fig. 4a). While the shape of the curve is somewhat chaotic and differs from run to run for the first few infill objective function evaluations, it quickly drops to zero suggesting that no infill sample point can be found to significantly improve the current best solution. Thus, we can terminate the iterations with confidence. We also observe that the points mapping predictions to observations neatly lie along the equal line, and that the standardized cross-validated residuals lie within the desirable interval $[-3,3]$ (see Fig. 4b). The sample points for constructing the RS are distributed widely across the search space (see Fig. 4c and d), manifesting the global exploration of the method. Later we show that RK without re-interpolation is prone to local exploitation. From the constructed RS, a narrow strip of the global optimal region can be identified centering around the abscissa (i.e. the toll rate for the first tolling interval) of 0.2 \$/km. Despite having random sampling and random search components, RK can produce consistent results across multiple runs.

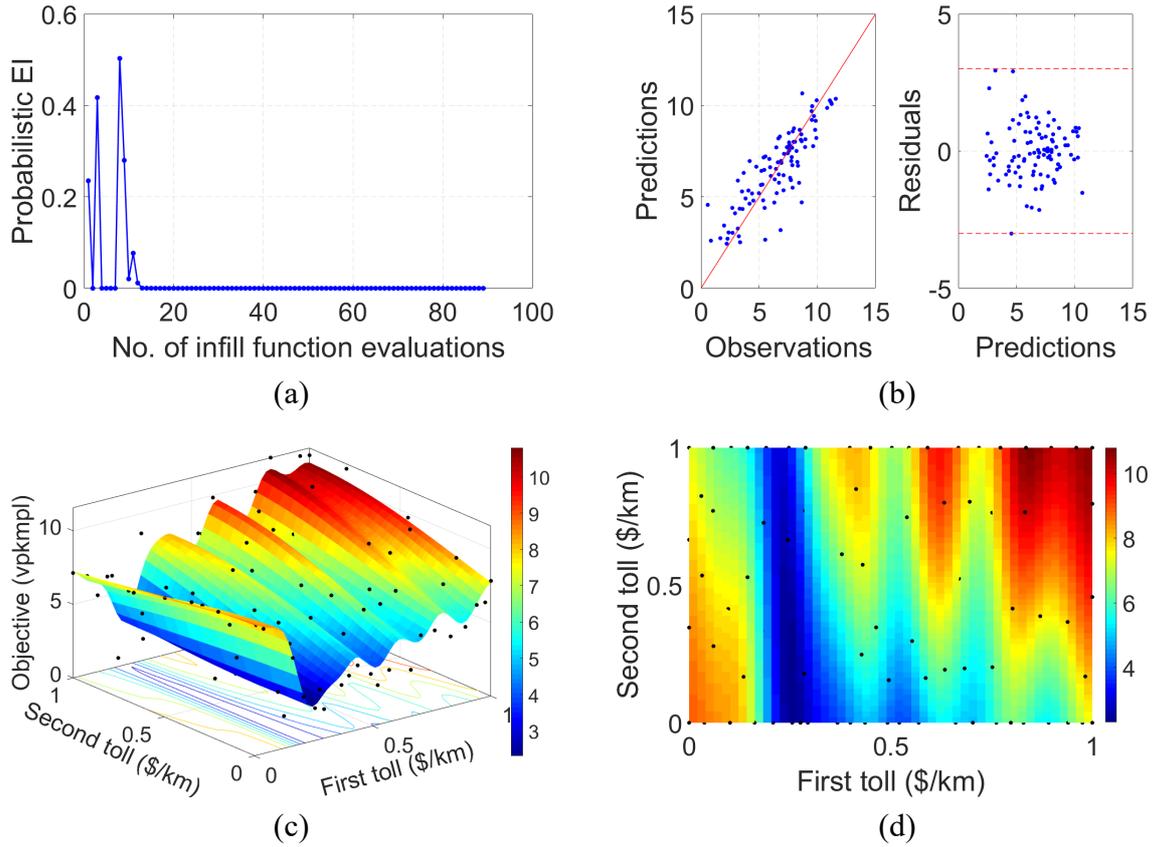

**Fig. 4.** Results of one run when applying RK to solve the simple problem: (a) convergence and (b) accuracy validation; and (c) and (d) constructed RS where the black dots are the sampled and evaluated points.

Fig. 5a shows that the optimal solutions obtained from all runs lie within the identified narrow strip of the global optimal region, as expected. Although the objective function values do not converge to the same optimum (see Fig. 5b), the differences are relatively small and mainly result from the numerical noise. This is demonstrated in Fig. 5c and d where an RS is constructed using sample points from all the ten runs (i.e. a total of 1,000 sample points) in conjunction interpolation[8] to allow for points that are not sampled. If one uses interpolation to construct the RS, overfitting is likely to occur as interpolating every single point is equivalent to capturing the numerical noise rather than filtering it out. Note that the objective function values reduce mostly within the first 30 objective function evaluations, which is consistent with the observation on the probabilistic EI metric (which converges to zero also within about 30 objective function evaluations).

---

[8] This is achieved using MATLAB's "griddata" function which performs triangular-based linear interpolation.

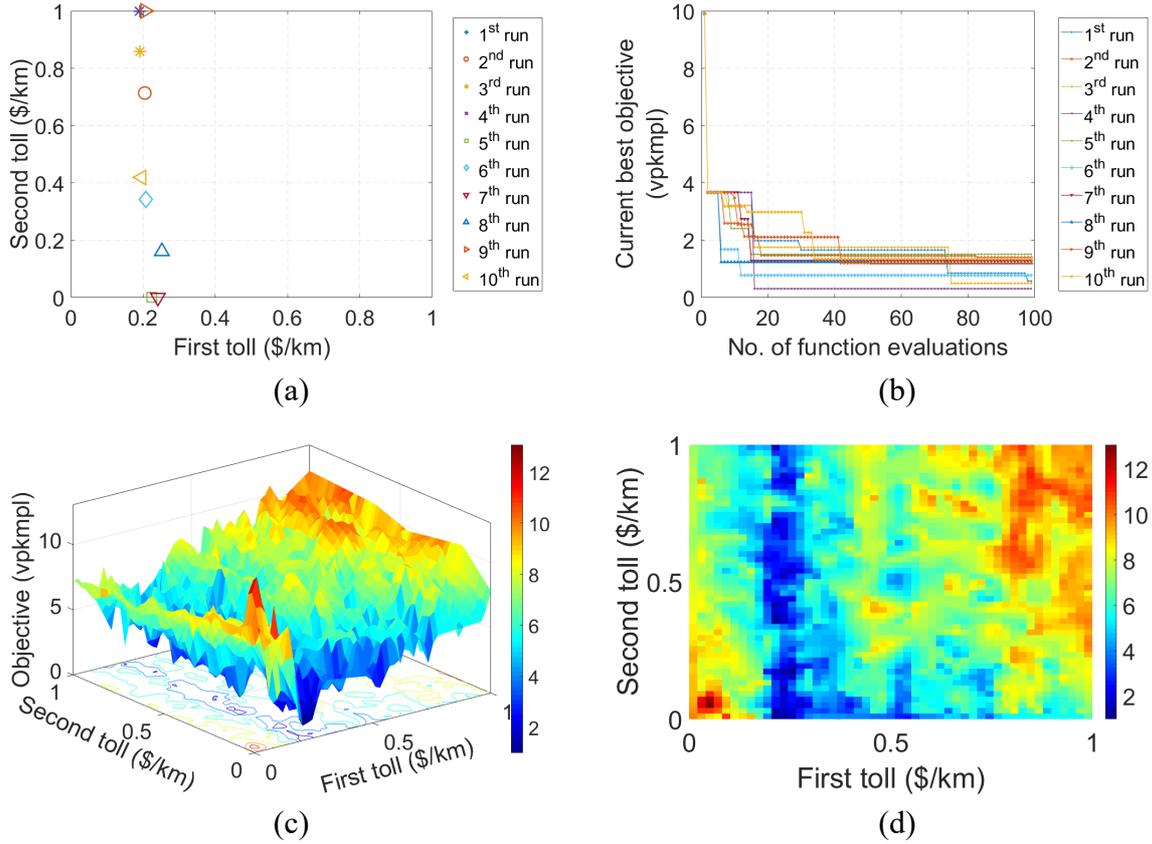

**Fig. 5.** Results of ten runs when applying RK to solve the simple problem: (a) spatial distribution of the optimal solutions; (b) evolution of the current best solution as the number of objective function evaluations increases; and (c) three- and (d) two-dimensional heatmaps of the interpolated RS using all the sample points.

We proceed to highlight the significance of re-interpolation that can help RK escape from local exploitation and retain global exploration. As shown in Fig. 6a and b, strong local exploitation appears in the absence of re-interpolation, as most sample points are concentrated or trapped in a small space that is only part of the identified narrow strip of the global optimal region. Due to underfitting, the constructed RS is highly biased that predicts poorly for the other parts of the search space. Fig. 6c and d show that, without re-interpolation, the probabilistic EI metric fluctuates unduly rather than converging to the ideal zero, and that the cross-validated residuals go beyond the desirable interval $[-3,3]$ for some predictions. If one is to predict for a few other points especially lying outside that small space, the predictions can significantly deviate from the true values.

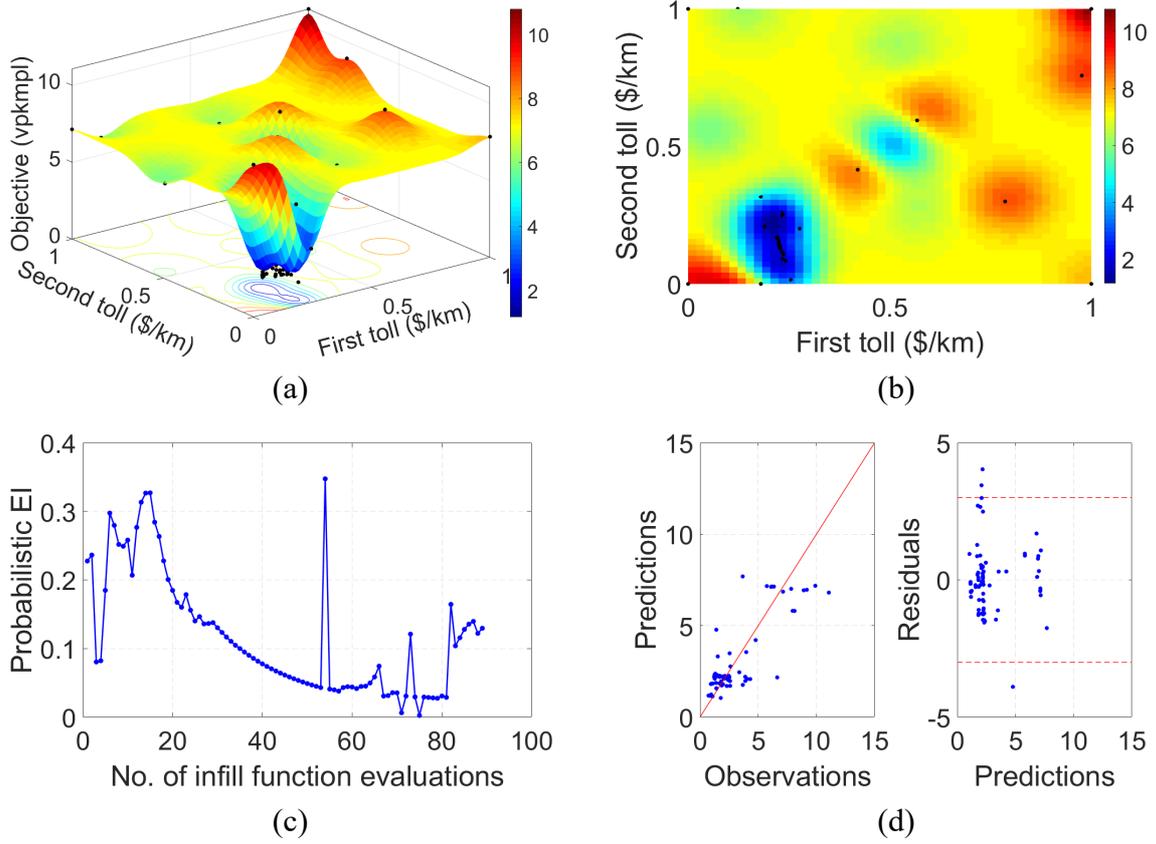

**Fig. 6.** Results obtained when applying RK without re-interpolation: (a) three- and (b) two-dimensional heatmaps of the constructed RS; and (c) convergence and (d) accuracy validation.

*5.1.3. DIviding RECTangles*

While DIRECT originates from Lipschitzian optimization, it shares a common characteristic with direct search – both require direct objective function evaluations only to guide the search. However, DIRECT is a global optimization method involving the least parameter tuning effort. The only user-specified parameter $\varepsilon$ turns out to exhibit limited effects on the performance of the method, and thus can be fixed to a value ranging from $10^{-7}$ to $10^{-3}$ (Jones et al., 1993).

Fig. 7 shows multiple contour plots of the objective function values as DIRECT proceeds with the iterations, sampling and evaluating more and more points. The final 35th iteration evaluates a total of 669 sample points, producing a result that highly resembles Fig. 4d. The optimal toll rates are 0.19 $/km for the first tolling interval and 0.95 $/km for the second, while the optimal objective function value is 0.005 vpkmpl (which is close to the ideal zero). Within about 100 objective function evaluations (see Fig. 7a), DIRECT can step into the global optimal region (specifically, the lower part of it), resulting in an optimal solution of $[0.24, 0.06]^T$ and an optimal objective function value of 0.874. The solution quality is about the same as that given by RK (see Fig. 5b). Note that a large proportion of the computational effort is spent on exploiting a local optimum in the lower middle part of the search space. From Fig. 7b and towards Fig. 7e, the global exploration characteristic of DIRECT becomes increasingly notable and the outline of the global optimal region gradually takes shape. Fig. 7f confirms that a good solution can be obtained within 100 objective function evaluations, despite not being the global optimum. With more objective function evaluations, the objective function value reduces further, as expected, and eventually converges to zero.

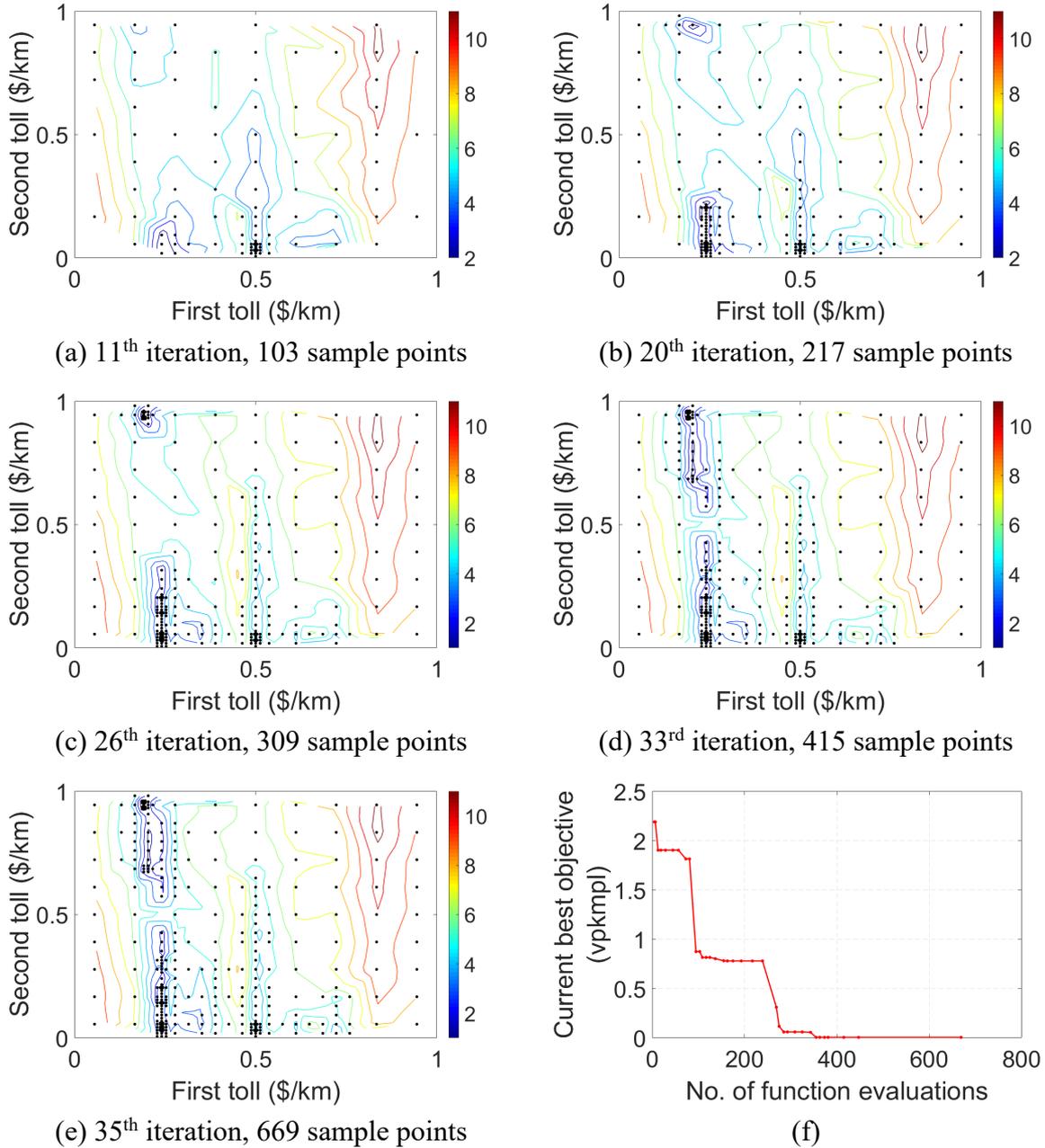

**Fig. 7.** Results obtained when applying DIRECT to solve the simple problem: (a)-(e) contour plots of the objective function values where the black dots are the sampled and evaluated points; and (f) evolution of the current best solution as the number of objective function evaluations increases.

### 5.1.4. Simultaneous perturbation stochastic approximation

SPSA requires careful parameter tuning especially in the presence of the numerical noise. There are in total five user-specified parameters along with their respective selection criteria (Spall, 1998). The parameter $c$ controls the accuracy of the gradient approximation and thus plays a critical role in the performance of the method – it should be set close to/away from zero if the numerical noise is low/high.

Fig. 8 shows multiple search paths of SPSA under six scenarios with varying $c$'s and initial points $\boldsymbol{\tau}_0$'s, while all the other parameters remain the same; i.e., $\alpha = 0.602$, $\gamma = 0.101$, $A = 5$, and $a = 0.1$. When (and only when) $c = 0.025$ and $\boldsymbol{\tau}_0 = [0.75, 0.75]^T$, SPSA is unable to search towards the narrow strip of the global optimal region (see Fig. 8f), which confirms with our previous discussion on how to choose $c$. As we have already observed the notable numerical noise in the objective function evaluations, $c$ should not be set close to zero so that the numerical noise can be filtered out, to some extent, during the gradient approximation. In doing so, the right search direction can be maintained. We have also observed that, apart from the narrow strip of the global optimal region, many local optima coexist resulting in a high chance of getting trapped in one of them. Indeed, $c = 0.025$ is not "strong" enough to provide the needed "bounce" for SPSA to escape from the local optimum and act as a global optimizer, not to mention that $\boldsymbol{\tau}_0 = [0.75, 0.75]^T$ is relatively far away from the global optimal region.

In contrast, when we set $c = 0.1$, SPSA quickly orients its search towards the global optimal region rather than wandering around the local optimum. The only concern is the resulting larger move per iteration, even though $a_i$, which controls the step size, is a decreasing function of the number of iterations. The reason is twofold: (i) a larger $c$ can result in a larger magnitude of the gradient approximation around the narrow strip of the global optimal region than in other parts of the search space; and (ii) the numerical noise can add to the magnitude of the gradient approximation. To make moves smaller and less aggressive, one can either use a smaller $a$ or scale down the gradient approximation (see Fig. 9a), both of which achieve the same effects of reducing the step size along the search path. However, there is no guarantee that the gradient approximation, or the gap between the objective function values of the two perturbed points, decays asymptotically to the ideal zero (see Appendix C). Even if one uses a small $c$ and the search path successfully reaches the global optimal region, the presence of the numerical noise prevents the two from vanishing.

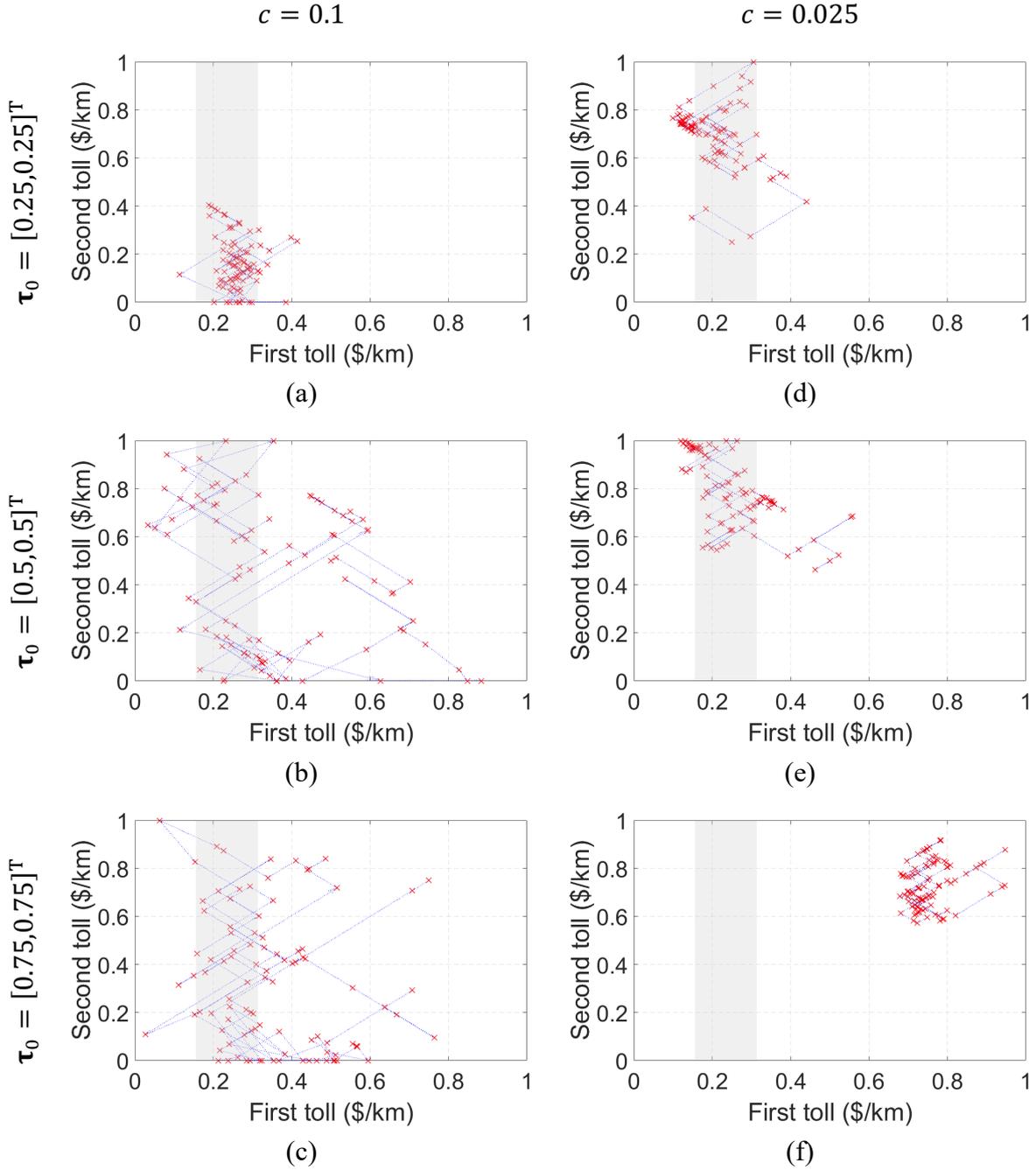

**Fig. 8.** Effects of different parameter settings on the performance of SPSA, where the red crosses are the calculated and evaluated points along the search path and the shaded area is the narrow strip of the global optimal region.

Note that SPSA only evaluates the two perturbed points at each iteration but not the point along the search path. Thus, to determine the optimal solution, one can either evaluate some additional points especially at the tail of the search path, or keep track of the current best solution from those perturbed points. Fig. 9b shows that, for different initial points, the objective function values reduce mostly within the first 50 objective function evaluations, which is a similar result to those of RK (see Fig. 5b) and DIRECT (see Fig. 7f).

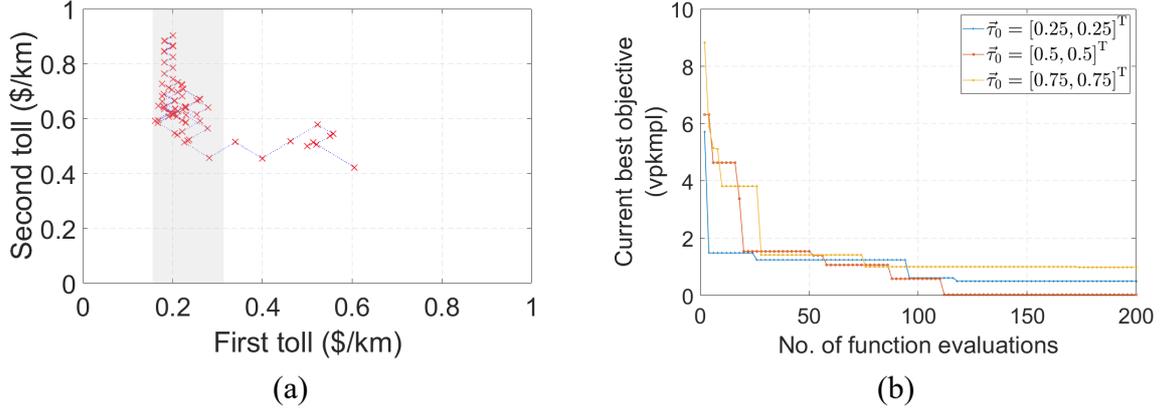

**Fig. 9.** (a) Search path of SPSA with $c = 0.1$ and $\boldsymbol{\tau}_0 = [0.5, 0.5]^T$, when the gradient approximation is scaled down; and (b) evolution of the current best solution for different initial points as the number of objective function evaluations increases.

## 5.2. Solving the complex problem

### 5.2.1. Regressing kriging

When applying RK to solve the complex TLP, we include the toll pattern smoothing constraints in the GA, thereby narrowing down the feasible region when searching for the optimal infill sample points. Considering 100 objective function evaluations as the computational budget and three runs to account for the randomness effect, Fig. 10a shows a gradually increasing trend, as expected, towards the end of the objective function evaluations (because we are now directly maximizing network flows throughout the tolling horizon). Fig. 10b shows the simulated NFDs of the PZ under the optimal tolling scenarios, where network flows maintain close to the maximum for a range of densities from 20 to 40 vpkmpl. We have already observed that, under the non-tolling scenario, $K_{cr}$ is somewhere between 20 and 30 vpkmpl (see Fig. 1c). Result here, however, suggests that $K_{cr}$ can evolve further to 40 vpkmpl in the presence of toll pricing, while maintaining rather high network flows. Given the objective of direct flow maximization, one need not struggle with identifying $K_{cr}$ to which the network shall be driven. Nor is it required to assume that the shape of the NFD including $K_{cr}$ does not change significantly when the network is pricing-controlled.

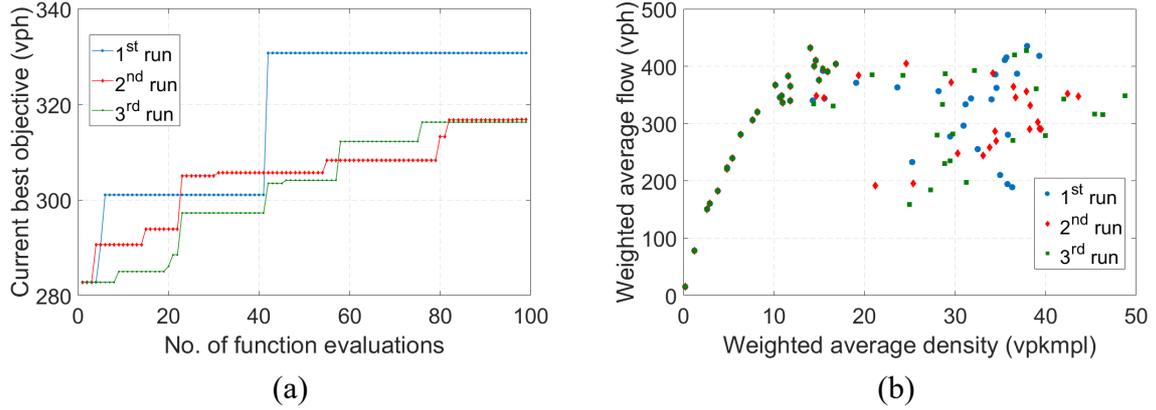

**Fig. 10.** Results obtained when applying RK to solve the complex problem with the objective of directly maximizing network flows: (a) evolution of the current best solution as the number of objective function evaluations increases; and (b) simulated NFDs of the PZ under the optimal tolling scenarios.

To demonstrate that the assumption does not always hold, i.e., the shape of the NFD including $K_{cr}$ may change in the presence of toll pricing, we further solve the same complex problem except with the indirect flow maximization objective (i.e. to operate the network around the identified $K_{cr}$ which is 25 vpkmpl). Fig. 11a shows that RK can quickly orient its search towards the global optimum within 100 objective function evaluations. The resulting NFDs, however, exhibit an unexpected and undesirable shape (see Fig. 11b). Although the network is well controlled around $K_{cr}$, the NFDs have already entered the congested regime resulting in rather low network flows. The new $K_{cr}$ turns out to be 15 vpkmpl, and the previous 25 vpkmpl no longer qualifies. Compared with the trapezoidal shape of the non-tolling NFD (see Fig. 1c), the new tolling NFDs have been somewhat squeezed to the left taking on a triangular shape. Thus, the optimal tolls are likely to be non-optimal given this change in the shape of the NFD. While one may increase $K_{cr}$ to be less conservative, it is still not guaranteed to obtain a similar shape of the NFD before and after pricing, and thus identifying $K_{cr}$ can become a tedious trial and error.

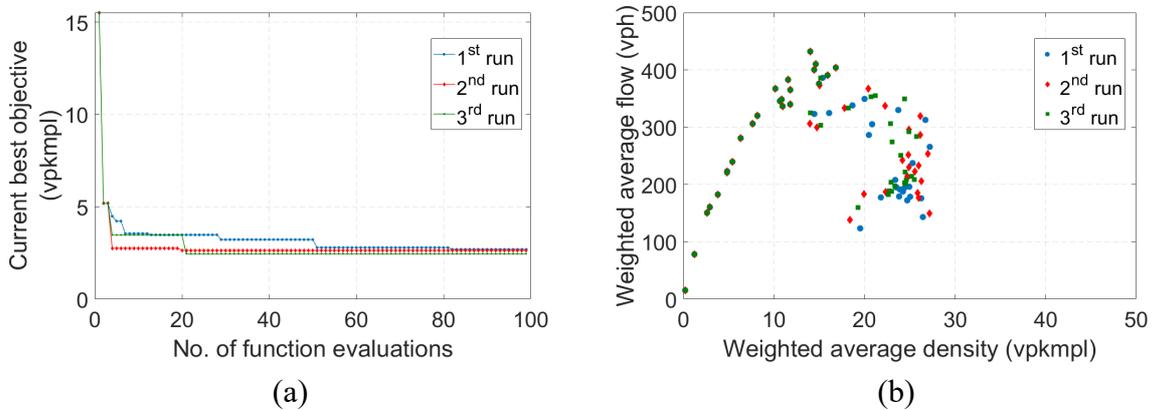

**Fig. 11.** Results obtained when applying RK to solve the complex problem with the objective of indirectly maximizing network flows using the critical network density: (a) evolution of the current best solution as the number of objective function evaluations increases; and (b) simulated NFDs of the PZ under the optimal tolling scenarios.

*5.2.2. The other two methods*

Due to the presence of the toll pattern smoothing constraints, DIRECT cannot be applied in its standard form to solve the complex TLP. Thus, we integrate DIRECT with the penalty function method (Bazaraa et al., 2013) to transform the original constrained optimization problem into an unconstrained one. In other words, if the toll pattern smoothing constraints are violated, a penalty is imposed on the objective function value, pushing DIRECT to search in the space where the constraints are met. A comparison between Fig. 12a and Fig. 10a suggests that DIRECT performs rather slowly in improving the current best solution to the complex problem. We have already observed that, within 100 objective function evaluations, RK can achieve an optimal objective function value of 320 vph. However, with the same number of objective function evaluations, DIRECT can only improve the objective function value to 285 vph. Even with nearly 600 objective function evaluations, the optimal objective function value is only slightly above 300 vph. One cause, as we previously discussed, is the existence of the numerical noise that cannot be tolerated by DIRECT. Another possible cause is the presence of the toll pattern smoothing constraints that limit the feasible region to a small part of the design space. As such, DIRECT may require a lot more objective function evaluations than RK given its iterative and exhaustive partitioning characteristic without directly accounting for the feasibility constraints.

When applying SPSA to solve the complex TLP, we similarly employ the penalty function method to handle the toll pattern smoothing constraints. For different initial points, we observe an expected increasing trend in the current best solution, as the number of objective function evaluations increases (see Fig. 12b). While a comparison between Fig. 12a and b suggests that SPSA can provide a slightly faster rate of convergence than DIRECT, different initial points chosen for the former can affect the search for the optimum (e.g., see the green curve in Fig. 12b). An ill-conditioned initial point can even trap the search at a bad local optimum, which we have observed already in Subsection 5.1.4.

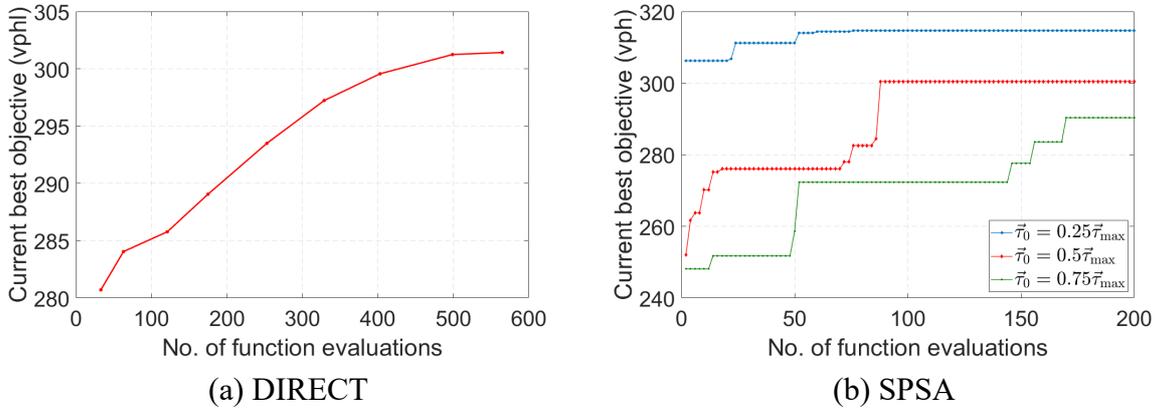

(a) DIRECT  (b) SPSA

**Fig. 12.** Evolution of the current best solution as the number of objective function evaluations increases, when applying both DIRECT and SPSA to solve the complex problem.

*5.3. Implications*

The four SBO methods have their own "intelligent" ways of leading the search towards the global optimum or optima, and thus are considered as representatives of computationally efficient methods among the big family of SBO (see Section 2). While this paper is all about toll pricing, the applicability of these methods is not limited to such problems; they can be applied to other transportation problems especially those replying on computer simulations. Our numerical experiments suggest that the four methods exhibit some performance differences when compared with one another, especially when applied to solve the complex problem. A few computational and practical insights are thus gained, but are no universal laws that apply to every conceivable transportation problem. Problem-specific features can be of help if one is interested in applying any of these methods to solve the problem at hand.

Specifically, results suggest that the four methods, when applied to solve the simple problem, perform equally well without significant difference. Recall that in this paper, the definition of a simple problem is one featuring a low dimensional decision vector, a set-point objective function, and only bound constraints. For such a simple problem, the PI controller performs particularly well resulting in the fastest rate of convergence according to our numerical experiments and comparison. Note that a strict mathematical proof of the convergence rate is impossible due to the existence of the "black-box" SBDTA model. The PI controller does require manual fine-tuning of the controller gain parameters, but studies have shown the robustness of the method to moderate parameter value changes (e.g., see Gu et al. (2018c)). Since the PI controller can only lead to a single global optimum, if multiple global optima coexist and one is more interested in finding all of them, then RK and DIRECT can be applied; see the heatmaps and contour plots in Subsections 5.1.2 and 5.1.3, respectively.

If a problem involves more decision variables (e.g. 16 vs. two in this paper), a complex objective function other than the set-point type (e.g. direct flow maximization), and/or complex constraints other than the bound type (e.g. toll pattern smoothing constraints), then RK, SPSA, and DIRECT are still applicable but not the PI controller. In other words, the scalability of the latter is far behind that of the former three. RK can handle and approximate a complex constraint by treating it as an independent RS, just like how it deals with the objective function. The infill sample points are thus selected not only based on how promising the objective function value will be (through the objective function RS), but also considering how likely the constraint will be met (through the constraint RS) (Forrester et al., 2008). SPSA and DIRECT can also handle a complex constraint, but in a quite different manner from RK. Specifically, the penalty function method is employed to transform the original constrained optimization problem into an unconstrained equivalent, which is then directly evaluated by the simulation (Bazaraa et al., 2013). Determining the penalty parameters, however, is usually non-trivial, which may require trial and error.

Results further suggest that RK, SPSA, and DIRECT exhibit quite different performance on the complex problem, especially in terms of the solution quality given the same computational budget (i.e. the same number of objective function evaluations). In short, RK turns out to be the best-performing method and DIRECT the least. As we previously discussed,

RK and DIRECT are the preferred methods for the simple problem if one strives to find all the global optima rather than a single optimum. In fact, even if the global optimum is unique, one can still apply these two methods to find the first few best solutions (analogous to the k-shortest paths). This is equivalent to providing multiple rather than a single choice for decision making. When the numerical noise (i.e. non-smoothness of the objective function) remains low, both methods work equally well. DIRECT can be a better alternative if parameter estimation required by RK for constructing the RS turns out to be far more time-consuming than the simulation itself; but this can only happen when the problem dimension is very high. We are not arguing, however, that DIRECT invariably outperforms RK for a very high dimensional problem, as this is a conjecture that requires further verification. Nevertheless, when both the numerical noise and the problem dimension increase, RK outperforms DIRECT irrespective of the problem dimension given its capability of filtering out the numerical noise and thus requiring fewer objective function evaluations to locate the global optimum or optima (see Fig. 10a and Fig. 12a). Even for the simple problem, DIRECT seems to waste some of the computational effort in exploiting a local optimum in the lower middle part of the search space (see Fig. 7).

Overall, RK is shown to improve the solution optimality at the fastest rate and achieve the best solution quality, thus being the most computationally efficient method for the complex problem. Due to the numerical noise as well as a higher problem dimension, the effectiveness and efficiency of SPSA and DIRECT reduce; the two methods can no longer compete with RK. Results suggest that the solution quality of SPSA is inferior to that of RK (see Fig. 10a and Fig. 12b), but is superior to that of DIRECT (see Fig. 12a and b). SPSA has the potential to act as a global optimizer for many challenging optimization problems, but no guarantee can be made that it never gets trapped in a local optimum. The possibility of the latter can be even greater if the numerical noise is high and one accidentally chooses an ill-conditioned initial point that is far away from the global optimum (see Fig. 8f and Fig. 12b). A straightforward solution is to inject Monte Carlo noise into the standard SPSA to avoid premature convergence, but this approach largely slows down the theoretical rate of convergence and requires greater effort for parameter tuning (Maryak and Chin, 2008). Our numerical experiments suggest that, compared with the other alternatives, the standard SPSA already requires the most effort for parameter tuning, which is a major concern if the objective function is computationally expensive (despite having basic guidelines in Spall (1998)). As with the PI controller, SPSA per run can only search (hopefully) for a single global optimum, but not every one of them if multiple coexist. A multi-start approach can be adopted at the cost of more computational time.

In summary, RK exhibits better applicability than the other alternatives. It proves to solve both the simple and complex problems more efficiently, even if the numerical noise is high. This result confirms and extends Huang et al. (2006) where several methods including RK were applied on deterministic test functions contaminated with noise. It may also explain the growing advocacy of RK in addressing simulation-based transportation problems (see Section 2). The PI controller, due to its limited applicability, is advisable only for a simple problem so as to provide a quick result. For a complex problem, DIRECT may perform badly due to the increased problem dimension and the presence of the numerical noise; SPSA can be a better alternative, but one must be aware of the potentially demanding requirement of parameter tuning and the risk resulting from an ill-conditioned initial point.

## 6. Conclusion

In this paper, we propose to solve two TLPs with different levels of complexity, where the NFD is used to indicate the level of congestion in the network. A large-scale SBDTA model of Melbourne, Australia is used in the problem formulation, resulting in an expensive-to-evaluate objective function. Thus, four state-of-the-art SBO methods are reviewed, applied, and compared as the solution algorithms. We find that all the methods work well in solving the simple problem. Although the PI controller is a more competitive solution to the simple problem given its faster rate of convergence, it is no longer applicable to the complex problem and thus is not a scalable method. RK exhibits the best overall performance especially for the complex problem, given its capability of filtering out the numerical noise arising from computer simulations. Note, however, that without re-interpolation, RK can lose the global exploration characteristic and only exploits a local region.

We emphasize that the chosen critical network density of the NFD does not necessarily represent a robust network control or optimization threshold, as it might shift in the presence of toll pricing. The conditions under which this shift can occur remain unknown, but we conjecture that a joint optimization approach accounting for both the tolling horizon and the toll levels can be useful in preventing such a shift as well as easing the significance of hysteresis. If one initiates toll pricing right after the network density exceeds the critical threshold, it might be too late for the price to drive the network back to its set-point state. In fact, to do so, the price can be rather high and thus disturbs and destabilizes the network (e.g. a large hysteresis loop of the NFD). This resembles the effect of a sharp drop in inflow to the network (Mahmassani et al., 2013). Thus, one can act in a proactive manner by treating the tolling horizon as part of the decision vector (i.e., each candidate tolling interval is a binary variable). Further research is required towards computationally efficient methods for solving the resulting mixed integer SBO problem.

We hope the results of this paper provide computational and practical insights into the application of the four SBO methods to toll pricing problems as well as other transportation problems. However, the results presented might depend on the case studies performed and thus shall be generalized further by testing on other large-scale networks. This is quite challenging as deploying a large-scale SBDTA model itself is a research task. Seeking existing models[9] are also difficult due to their low transferability and accessibility (e.g. different (commercialized) platforms and languages). Thus, we urge a concerted effort in this regard.

**Acknowledgments**

The authors would like to thank the four anonymous reviewers for their constructive comments and suggestions which significantly improve the quality of the paper.

---

[9] See Shafiei et al. (2018) for a brief summary of some of these models.

# Appendix A. Algorithmic details of the four SBO methods

Here, we describe the algorithmic details of the four SBO methods. For better readability, a summary of notations is provided in Table A. 1.

**Table A. 1**
Summary of notations for the four SBO methods.

| Method | Notation | Interpretation |
|---|---|---|
| PI controller | $\tau_h(i)$ | Toll rate for the $h$th tolling interval at iteration $i$ |
| | $P_P/P_I$ | Proportional/integral gain parameter |
| | $N_{\max}$ | Maximum number of iterations allowed |
| RK | $y(\cdot)$ | True response surface |
| | $\mu/\hat{\mu}$ | Unknown constant mean of the true response surface/maximum likelihood estimate (MLE) |
| | $\sigma^2/\hat{\sigma}^2$ | Process variance/MLE |
| | $\psi(\cdot)$ | Gaussian correlation function |
| | $\theta_l$ | $l$th entry of the scaling coefficients vector |
| | $\lambda/\hat{\lambda}$ | Regularization constant/MLE |
| | $\mathbf{R}$ | Regressing correlation matrix |
| | $\boldsymbol{\psi}$ | Correlation vector between the new point and all the existing points |
| | $\hat{y}(\cdot)/\hat{s}^2(\cdot)$ | RK predictor/prediction error |
| | $y_{\min}$ | Current best solution (for a minimization problem) |
| DIRECT | $\mathbf{c}_j$ | Midpoint of the $j$th hyperrectangle |
| | $d_j$ | Distance between $\mathbf{c}_j$ and the vertices of the $j$th hyperrectangle |
| | $\mathcal{D}^\hbar$ | Set of dimensions with the longest side length of the potentially optimal hyperrectangle $\hbar$ |
| | $\delta^\hbar$ | One third of the length of $\mathcal{D}^\hbar$ |
| | $\mathbf{c}^\hbar$ | Midpoint of the potentially optimal hyperrectangle $\hbar$ |
| | $\delta^\hbar$ | One third of the longest side length |
| | $\mathbf{e}_i$ | Unit vector along the $i$th dimension of $\mathcal{D}^\hbar$ |
| SPSA | $\Delta_{il}$ | $l$th entry of the random perturbation vector at iteration $i$ |
| | $a_i/c_i$ | Gain sequences at iteration $i$ |
| | $\hat{\mathbf{g}}_i(\cdot)$ | Gradient approximation at iteration $i$ |

*A.1. Proportional-integral controller*

The mathematical formulation of the discrete PI controller for the iteration counter $i = 1$ is

$$\tau_h(i) = P_I \cdot (\bar{K}_h(i) - K_{cr}), \ h = 1,2,\ldots,m, \tag{A.1}$$

where $\tau_h(i)$ is the adjusted toll rate for the $h$th tolling interval during iteration $i$ and $P_I > 0$ is the integral gain parameter. When $i > 1$,

$$\tau_h(i) = \tau_h(i-1) + P_P \cdot \left(\bar{K}_h(i) - \bar{K}_h(i-1)\right) + P_I \cdot (\bar{K}_h(i) - K_{cr}),$$
$$h = 1, 2, \ldots, m, \tag{A.2}$$

where $P_P > 0$ is the proportional gain parameter. If a mathematical description of the system dynamics is available, one can analytically derive $P_P$ and $P_I$ (Keyvan-Ekbatani et al., 2012; Keyvan-Ekbatani et al., 2015). But because we use a "black-box" SBDTA model, these parameters can only be estimated via trial and error (Gu et al., 2018c) – one can pick some values aggressively at the beginning and gradually reduce them to ease the observed oscillations. Eqs. (A.1) and (A.2) are incorporated into the following framework:

**Step 1.** Run the simulation without pricing to get the non-tolling NFD of the PZ.
**Step 2.** Set $i = 1$ and calculate the initial toll rates via Eq. (A.1).
**Step 3.** Run the simulation with the newly calculated toll rates to get the updated NFD of the PZ; set $i = i + 1$.
**Step 4.** If $i \leq N_{max}$ where $N_{max}$ is the maximum number of iterations allowed, calculate the updated toll rates via Eq. (A.2) and go back to Step 3; otherwise, terminate the algorithm.

*A.2. Regressing kriging*

The workflow of RK is illustrated in Fig. A. 1 consisting of three major components. These components are described in detail below.

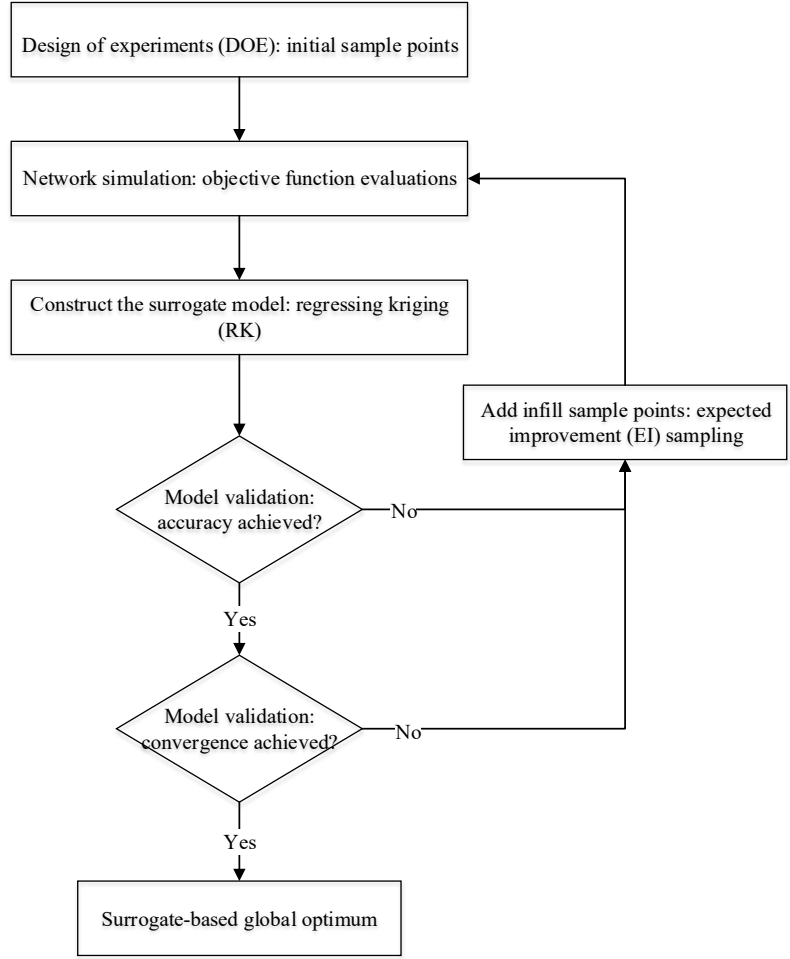

**Fig. A. 1.** Workflow of RK.

*A.2.1. Experiments design*

Experiments design, or design of experiments, is intended to provide an initial set of sample points so as to construct the starting surrogate model. The space-filling Latin Hypercube Sampling (LHS) is one such method that spreads the initial sample points as uniformly as possible over the entire feasible domain. Since each problem dimension is stratified into an equal number of intervals from which points are uniformly sampled, there is no overlap when mapping the multi-dimensional data points into each dimension. To spread the initial sample points to the greatest extent, the maximin LHS can be applied (Forrester et al., 2008).

*A.2.2. Response surface construction*

Ordinary kriging assumes an unknown constant mean $\mu$ of the true RS $y(\boldsymbol{\tau})$, plus a set of basis functions centered around the sample points. The covariance between any two points is $\sigma^2 \psi(\boldsymbol{\tau}^{(i)}, \boldsymbol{\tau}^{(j)})$, where $\sigma^2$ is the process variance and $\psi(\boldsymbol{\tau}^{(i)}, \boldsymbol{\tau}^{(j)}) = \exp\left(-\sum_{l=1}^{m} \theta_l \cdot \left(\tau_l^{(i)} - \tau_l^{(j)}\right)^2\right)$ is the Gaussian kernel depending on the distance between the two points only ($\boldsymbol{\theta}$ is a vector of scaling coefficients). As an interpolation method (i.e., the constructed RS passes through all the sample points), ordinary kriging is not advisable when computer simulations exhibit high numerical noise, because then interpolation can lead to overfitting of the RS (Forrester et al., 2006). The remedy is to regress rather than interpolate the sample points, which can be achieved by introducing a regularization constant $\lambda$ to the diagonal of the correlation matrix, resulting in the regressing correlation matrix $\mathbf{R}$. By maximizing the augmented log likelihood using a Gaussian distribution (Forrester et al., 2006), the RK predictor and the associated prediction error can be obtained:

$$\hat{y}(\boldsymbol{\tau}^*) = \hat{\mu} + \boldsymbol{\psi}^T \mathbf{R}^{-1}(\mathbf{y} - \mathbf{1}\hat{\mu}), \tag{A.3}$$

$$\hat{s}^2(\boldsymbol{\tau}^*) = \hat{\sigma}^2\big(1 + \hat{\lambda} - \boldsymbol{\psi}^T \mathbf{R}^{-1} \boldsymbol{\psi}\big), \tag{A.4}$$

where $\hat{\mu}$, $\hat{\sigma}^2$, and $\hat{\lambda}$ are the MLEs, and $\boldsymbol{\psi}$ is the correlation vector between the new point $\boldsymbol{\tau}^*$ and all the existing sample points. Clearly, the prediction is treated as a distribution with a mean and a variance. The confidence of the prediction is high if the variance is low, and vice versa.

To enhance the constructed RS, infill sample points are added via EI sampling, which is an infill strategy that can balance well between exploring unvisited regions and exploiting visited regions (Chen et al., 2014; Ekström et al., 2016). While seeking to locate an infill sample point that leads to a better prediction, EI sampling also considers the confidence of the prediction – in a region with few sample points and poor predictions, the uncertainty is high suggesting a good opportunity to improve the current best solution by exploring this region. Denoting the current best solution by $y_{\min}$, the EI at an infill sample point reads $\mathbb{E}[I(\boldsymbol{\tau}^*)] = \mathbb{E}[\max(y_{\min} - y(\boldsymbol{\tau}^*), 0)]$. When $\hat{s}^2(\boldsymbol{\tau}^*) = 0$, $\mathbb{E}[I(\boldsymbol{\tau}^*)] = 0$; when $\hat{s}^2(\boldsymbol{\tau}^*) > 0$,

$$\mathbb{E}[I(\boldsymbol{\tau}^*)] = \frac{1}{\sqrt{2\pi}\hat{s}^2(\boldsymbol{\tau}^*)} \int_{-\infty}^{y_{\min}} (y_{\min} - u) \exp\left(-\frac{(u-\hat{y}(\boldsymbol{\tau}^*))^2}{2\hat{s}^2(\boldsymbol{\tau}^*)}\right) du. \tag{A.5}$$

To prevent RK from getting trapped at an existing sample point, Forrester et al. (2006) proposed a re-interpolation MLE of $\sigma^2$:

$$\hat{\sigma}_{\text{ri}}^2 = \frac{(\mathbf{y}-\mathbf{1}\hat{\mu})^T \mathbf{R}^{-1} \boldsymbol{\Psi} \mathbf{R}^{-1}(\mathbf{y}-\mathbf{1}\hat{\mu})}{n}. \tag{A.6}$$

The associated re-interpolation prediction error reads $\hat{s}_{\text{ri}}^2(\boldsymbol{\tau}^*) = \hat{\sigma}_{\text{ri}}^2(1 - \boldsymbol{\psi}^T \mathbf{R}^{-1} \boldsymbol{\psi})$.

*A.2.3. Model validation*

To validate RK, the leave-one-out cross validation (CV) can be applied that does not require sampling and evaluating any additional sample points. With a total of $n$ observations, the leave-one-out CV is repeated $n$ times resulting in a cross-validated prediction $\hat{y}_{-i}(\mathbf{x}^{(i)})$ and a standard error $\hat{s}_{-i}(\mathbf{x}^{(i)})$ corresponding to the observed $y(\mathbf{x}^{(i)})$. One can thus calculate the 99.7 percent confidence interval for each $y(\mathbf{x}^{(i)})$ using the prediction plus or minus three standard errors (Jones et al., 1998). If one calculates the standardized cross-validated residual $\frac{y(\mathbf{x}^{(i)}) - \hat{y}_{-i}(\mathbf{x}^{(i)})}{\hat{s}_{-i}(\mathbf{x}^{(i)})}$, the value should ideally be lying within the interval $[-3,3]$.

*A.3. DIviding RECTangles*

- **Step 1.** *Initialization.* Normalize the search space into a unit hypercube whose midpoint is denoted by $\mathbf{c}_1$. Run the simulation and evaluate the objective function value $y(\mathbf{c}_1)$. For a minimization problem, set the current best solution $y_{\min} = y(\mathbf{c}_1)$, and set the iteration counter $i = 1$.
- **Step 2.** *Potentially optimal hyperrectangle identification.* Suppose the search space is currently partitioned into $k$ hyperrectangles. We use $\mathbf{c}_j$ to denote the midpoint of the $j$ th hyperrectangle and $d_j$ to denote the distance between $\mathbf{c}_j$ and the hyperrectangle vertices. A hyperrectangle $j^*$ is potentially optimal if the following two inequalities hold for some $\widetilde{K} > 0$, where $\varepsilon > 0$ is a small constant:

$$y(\mathbf{c}_{j^*}) - \widetilde{K} d_{j^*} \leq y(\mathbf{c}_j) - \widetilde{K} d_j, \ j \in (1,2,\dots,k), \tag{A.7}$$

$$y(\mathbf{c}_{j^*}) - \widetilde{K} d_{j^*} \leq y_{\min} - \varepsilon |y_{\min}|. \tag{A.8}$$

- **Step 3.** *Hyperrectangle partitioning.* Let $\mathcal{H}$ denote the set of potentially optimal hyperrectangles. For each $\hbar \in \mathcal{H}$, denote the set of dimensions with the longest side length by $\mathcal{D}^\hbar$ and set $\delta^\hbar$ at one third of this length. Sample and evaluate the objective function values at points $\mathbf{c}^\hbar \pm \delta^\hbar \mathbf{e}_l$, where $\mathbf{c}^\hbar$ is the midpoint of $\hbar$ and $\mathbf{e}_l$ is the unit vector along the $l$ th dimension in $\mathcal{D}^\hbar$. Calculate $w_l = \min\left(y(\mathbf{c}^\hbar + \delta^\hbar \mathbf{e}_l), y(\mathbf{c}^\hbar - \delta^\hbar \mathbf{e}_l)\right)$ and partition $\hbar$ into thirds along each dimension $l \in \mathcal{D}^\hbar$ in ascending order of $w_l$, i.e. starting with the dimension having the smallest $w_l$ and continuing to the dimension having the largest $w_l$.
- **Step 4.** *Stop test.* Update $y_{\min}$. Terminate the algorithm if the maximum number of iterations allowed is reached; otherwise, set $i = i + 1$ and go back to Step 2.

*A.4. Simultaneous perturbation stochastic approximation*

Given an $m$-dimensional minimization problem, SPSA generates, at each iteration, an $m$-dimensional random perturbation vector $\mathbf{\Delta}_i$. Each element $\Delta_{il}$, where $l \in (1,2,\dots,m)$, is independently generated by Monte Carlo from any zero-mean probability distributions satisfying the SPSA regularity conditions (Spall, 1992). One such distribution commonly used is the Bernoulli distribution with 0.5 probability for $\pm 1$, which results in the following

approximated gradient:

$$\hat{\mathbf{g}}_i(\boldsymbol{\tau}_i) = \frac{y(\boldsymbol{\tau}_i+c_i\boldsymbol{\Delta}_i)-y(\boldsymbol{\tau}_i-c_i\boldsymbol{\Delta}_i)}{2c_i}\begin{bmatrix}\Delta_{i1}^{-1}\\\vdots\\\Delta_{ik}^{-1}\end{bmatrix},\quad (A.9)$$

where $c_i = \frac{c}{(i+1)^\gamma}$ is a small number for the gradient approximation with user-specified parameters $c$ and $\gamma$. A practically effective and theoretically valid value of $\gamma$ is 0.101, while $c$ must not be set close to zero given a noisy objective function (Spall, 1998). The detailed algorithmic steps are as follows:

**Step 1.** *Parameter initialization.* Set $i = 1$ and initialize $\boldsymbol{\tau}_1 \in \mathbb{R}^m$. Set $c_i$ and the step size $a_i = \frac{a}{(A+i)^\alpha}$, where $a$, $A$, and $\alpha$ are user-specified parameters. A complete guideline for choosing such parameters can be found in Spall (1998).

**Step 2.** *Random perturbation.* Generate a $m$-dimensional random perturbation vector $\boldsymbol{\Delta}_i$, where each element is independently sampled via Monte Carlo from a Bernoulli distribution with 0.5 probability for $\pm 1$.

**Step 3.** *Objective function evaluation.* Run the simulation for both the perturbed decision vectors and evaluate their objective function values.

**Step 4.** *Gradient approximation.* Calculate the approximated gradient using Eq. (A.9).

**Step 5.** *Decision vector update.* Apply the following standard SA formulation (for a minimization problem) to update the decision vector:

$$\boldsymbol{\tau}_{i+1} = \boldsymbol{\tau}_i - a_i\hat{\mathbf{g}}_i(\boldsymbol{\tau}_i). \quad (A.10)$$

**Step 6.** *Stop test.* Terminate the algorithm if there is little change in several successive gradient approximations or objective function evaluations, or the maximum number of iterations allowed is reached; otherwise, set $i = i + 1$ and go back to Step 2.

# Appendix B. Results for multiple runs of regressing kriging

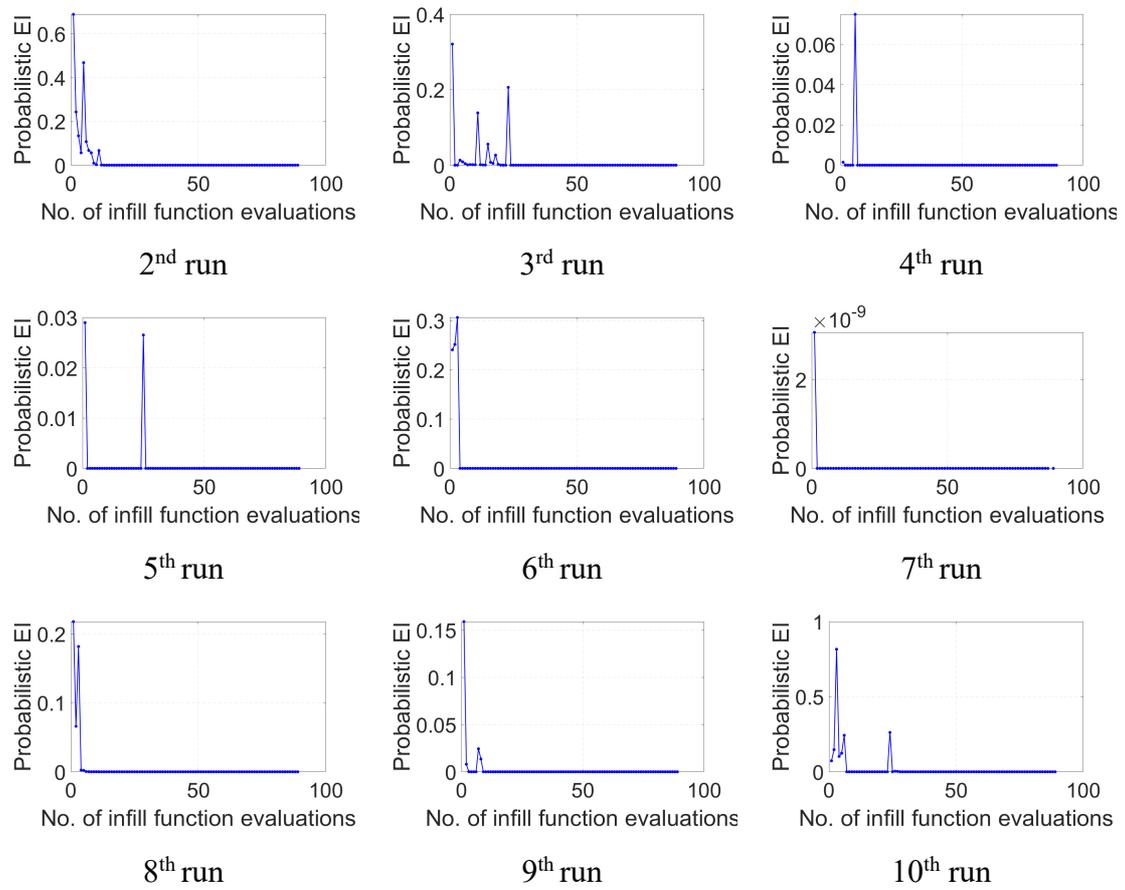

**Fig. B. 1.** Validating convergence using the probabilistic EI metric for multiple runs of RK.

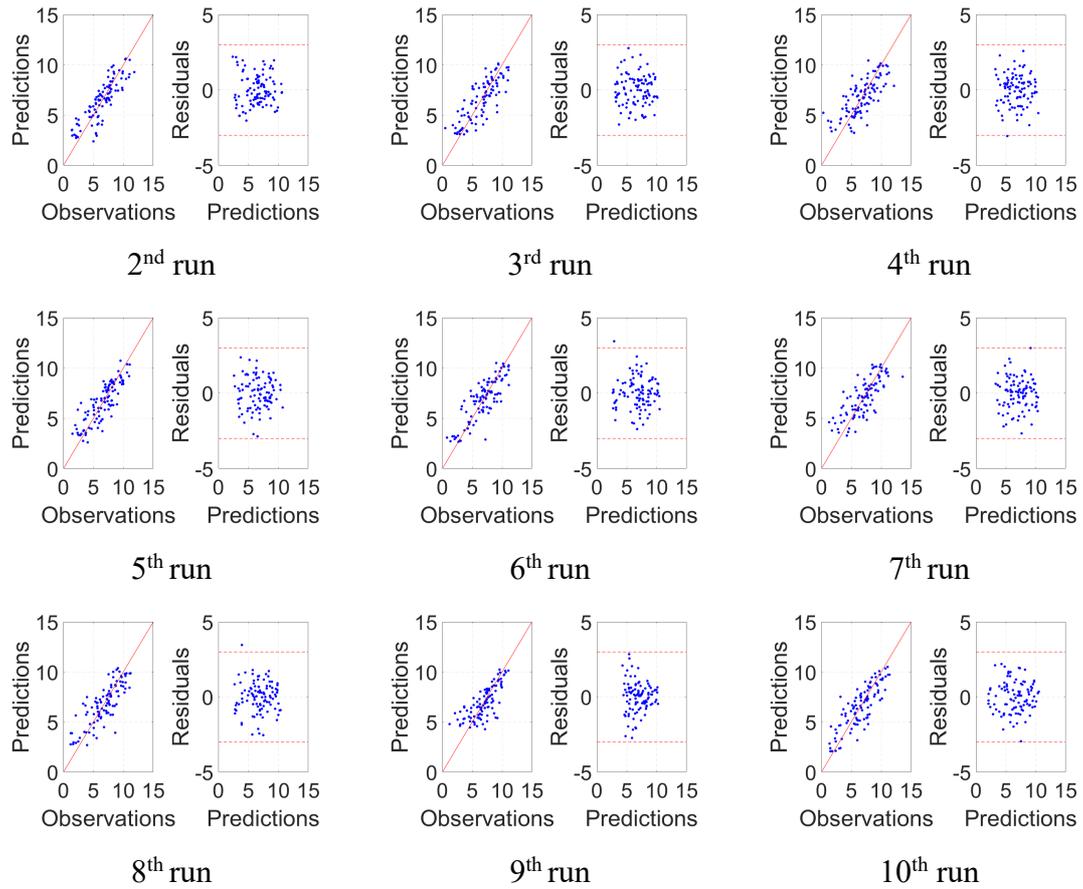

**Fig. B. 2.** Validating accuracy using the standardized cross-validated residuals for multiple runs of RK.

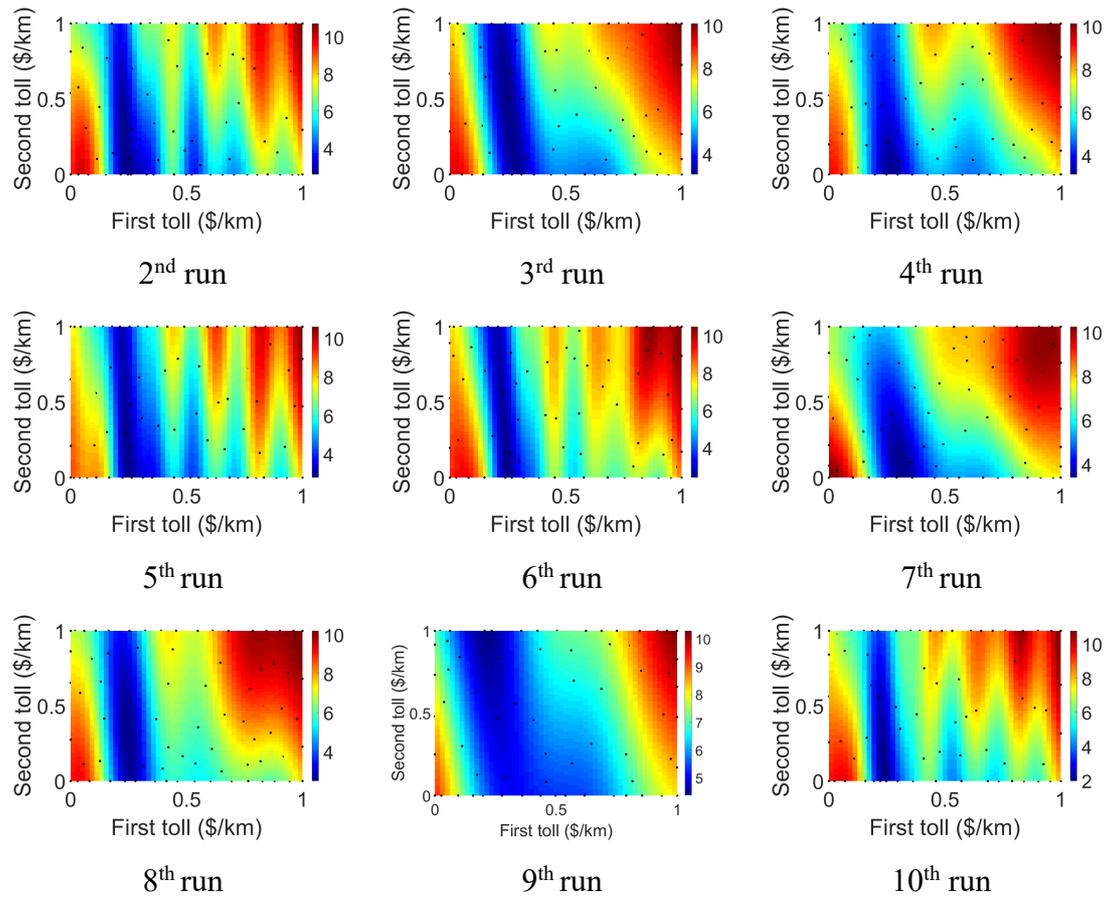

**Fig. B. 3.** Constructed RSs for multiple runs of RK where the black dots are the sampled and evaluated points.

**Appendix C. Results for simultaneous perturbation stochastic approximation**

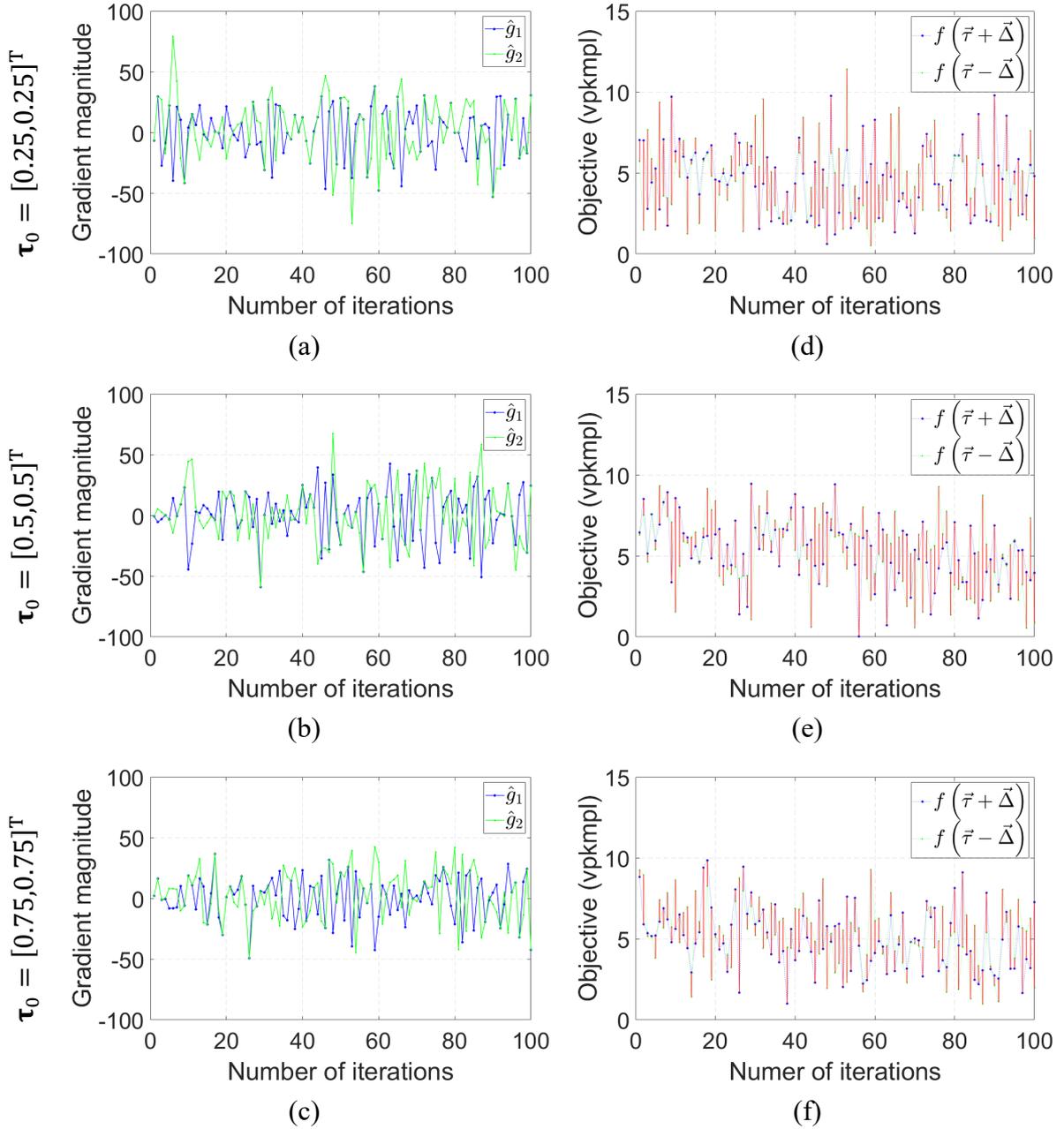

**Fig. C. 1.** (a)-(c) Magnitude of the gradient approximation in SPSA considering different initial points; and (d)-(f) gaps between the objective function values of the two perturbed points at every iteration of SPSA using $c = 0.1$, considering different initial points.

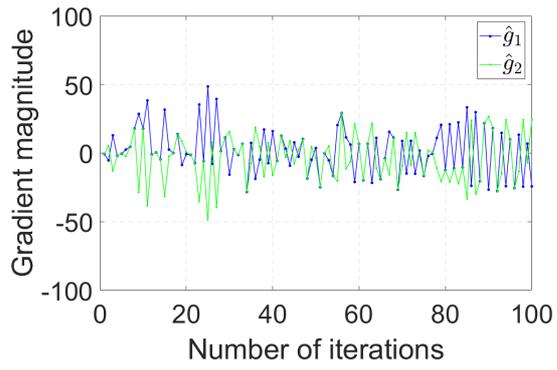 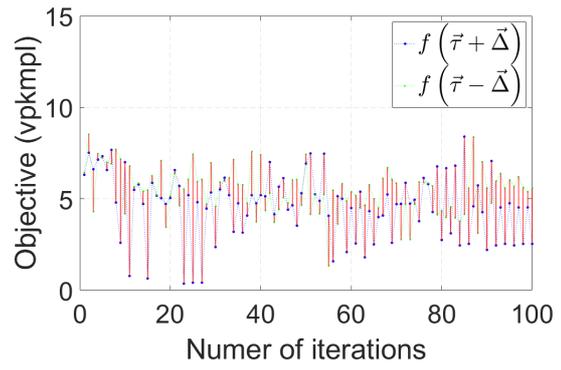

(a)                  (b)

**Fig. C. 2.** Effects of scaling down the gradient approximation in SPSA using $c = 0.1$ and $\boldsymbol{\tau}_0 = [0.5, 0.5]^{\mathrm{T}}$: (a) magnitude of the gradient approximation; and (b) gaps between the objective function values of the two perturbed points at every iteration.